\documentclass[12pt,eqno,epsf]{article}
\usepackage{amsmath,amssymb,graphicx,here}
\numberwithin{equation}{section}

\def\ignore#1{{}}

\newcounter{sxn}

\newcounter{axn}

\date{}

\newdimen\mybaselineskip
\renewcommand{\thefootnote}{\arabic{footnote}}

\newcommand{\beeq}{\begin{equation}}
\newcommand{\eneq}{\end{equation}}
\newcommand{\beqn}{\begin{eqnarray}}
\newcommand{\eeqn}{\end{eqnarray}}

\newcommand{\tht}{\theta}

\newcommand{\kp}{\kappa}

\newcommand{\sgm}{\sigma}

\newcommand{\vph}{\varphi}

\newcommand{\be}{\begin{equation}}
\newcommand{\ee}{\end{equation}}
\newcommand{\bea}{\begin{eqnarray}}
\newcommand{\eea}{\end{eqnarray}}
\newcommand{\eql}{\!\!\!&=\!\!\!&}

\newcommand{\defa}{\!\!\!&\equiv\!\!\!&}

\newcommand{\toa}{\!\!\!&\to\!\!\!&}

\newcommand{\tl}[1]{\tilde{#1}}
\newcommand{\bdm}[1]{{\mbox{\boldmath $#1$}}}

\newcommand{\der}{\partial}

\newcommand{\ie}{{i.e.}}

\newcommand{\vev}[1]{\langle #1 \rangle}

\newcommand{\brkt}[1]{\left( #1 \right)}
\newcommand{\brc}[1]{\left\{ #1 \right\}}
\newcommand{\sbk}[1]{\left[ #1 \right]}
\newcommand{\abs}[1]{\left| #1 \right|}

\newcommand{\cG}{{\cal G}}

\newcommand{\cL}{{\cal L}}

\newcommand{\cR}{{\cal R}}

\begin{document}
\thispagestyle{empty}

\baselineskip=12pt


\begin{flushright}
KEK-TH-2259 \\
WU-HEP-20-09
\end{flushright}

\baselineskip=25pt plus 1pt minus 1pt

\vskip 1.5cm

\begin{center}
{\LARGE\bf Fate of domain walls in 5D gravitational theory with compact extra dimension}

\vspace{1.0cm}
\baselineskip=20pt plus 1pt minus 1pt

\normalsize

{\large\bf Hiroyuki Abe,}${}^1\!${\def\thefootnote{\fnsymbol{footnote}}
\footnote[1]{E-mail address: abe@waseda.jp}}
{\large\bf Shuntaro Aoki}${}^1\!${\def\thefootnote{\fnsymbol{footnote}}
\footnote[2]{E-mail address: shun-soccer@akane.waseda.jp}}
{\large\bf Yu Asai}${}^1\!${\def\thefootnote{\fnsymbol{footnote}}
\footnote[3]{E-mail address: u-asai.physics@ruri.waseda.jp}} \\
{\large\bf and Yutaka Sakamura}${}^{2,3}\!${\def\thefootnote{\fnsymbol{footnote}}
\footnote[4]{E-mail address: sakamura@post.kek.jp}}

\vskip 1.0em

${}^1${\small\it Department of Physics, Waseda University, \\ 
3-4-1 Ookubo, Shinjuku-ku, Tokyo 169-8555, Japan}

\vskip 1.0em

${}^2${\small\it KEK Theory Center, Institute of Particle and Nuclear Studies, 
KEK, \\ 1-1 Oho, Tsukuba, Ibaraki 305-0801, Japan} \\ \vspace{1mm}
${}^3${\small\it Department of Particles and Nuclear Physics, \\
SOKENDAI (The Graduate University for Advanced Studies), \\
1-1 Oho, Tsukuba, Ibaraki 305-0801, Japan}

\end{center}

\vskip 1.0cm
\baselineskip=20pt plus 1pt minus 1pt

\begin{abstract}
We pursue the time evolution of the domain walls in 5D gravitational theory with a compact extra dimension 
by numerical calculation. 
In order to avoid a kink-antikink pair that decays into the vacuum, we introduce a topological winding 
in the field space. 
In contrast to the case of non-gravitational theories, there is no static domain-wall solution in the setup. 
In the case that the minimal value of the potential is non-negative, 
we find that both the 3D space and the extra dimension will expand at late times  
if the initial value of the Hubble parameter is chosen as positive. 
The wall width almost remains constant during the evolution. 
In other cases, the extra dimension diverges and the 3D space shrinks to zero at a finite time. 
\end{abstract}

\newpage

\section{Introduction}
The possibility that our four-dimensional (4D) spacetime is localized on a domain wall 
in the extra dimension has been extensively investigated as one of the simplest setups 
for the braneworld scenario~\cite{Rubakov:1983bb}-\cite{Okada:2019fgm}. 
Most of them considered the infinite extra dimension. 
This is because a kink configuration generically induces an antikink configuration 
due to the periodic boundary condition along the extra dimension, 
and such a field configuration is unstable and decays into the vacuum. 
However, this is not the case when the field space is compact and the gravity is neglected. 
In such a case, a stable kink solution can exist. 
The stability is ensured by the topological winding around the compact field space~\cite{Maru:2001gf,Eto:2004zc}. 

Such a compact field space appears in various effective theories 
as a phase of a complex scalar field that has 
a nonvanishing vacuum expectation value, just like the axion. 
Hence the possibility that we live on a domain wall in such theories is worthwhile to consider. 
Although there are a vast amount of the braneworld models in compact extra dimensions, 
few of them discuss the effects of the finite width of the brane 
by treating the brane as a field-theoretical soliton. 
Here, we consider domain walls in the compact extra dimension, and investigate how their field configuration  
affects the time evolutions of the three-dimensional (3D) space and the extra dimension. 

In order to discuss the cosmological evolution of the braneworld scenario, 
the gravity must be taken into account. 
When the gravity is turned on, the situation for the stability changes. 
The positive tension of the domain wall warps the ambient geometry,  
just like the Randall-Sundrum model~\cite{Randall:1999vf}. 
Since each domain wall decreases the derivative of the warp factor~\cite{Hatanaka:1999ac}, 
the periodic boundary condition for the warp factor cannot be satisfied. 
This indicates that there is no static domain-wall solution in the gravitational theory 
with the compact extra dimension. 
Still, we can introduce the topological winding around the field space even in such theories. 
In this paper, we consider a field configuration with nonzero winding number, 
and pursue its time evolution by numerical calculation. 

The paper is organized as follows. 
In the next section, we briefly review the case of non-gravitational theory, 
and provide an analytic expression for a static domain-wall solution 
in the compact extra dimension. 
In Sec.~\ref{DWinG}, we extend it to the gravitational theory, 
and give the field equations to solve. 
In Sec.~\ref{num_result}, we show the numerical results for the time evolution 
of the domain walls. 
Sec.~\ref{summary} is devoted to the summary. 
In Appendix~\ref{Ja}, we collect the definition and some properties of the Jacobi amplitude, 
which expresses the initial domain-wall configuration. 
In Appedix~\ref{static_limit_EOM}, we provide a direct relation between the metric ansatze 
used in Secs.~\ref{DWinG:1} and \ref{DWinG:2}.

\section{Case of non-gravitational theory}
\subsection{Topological winding in field space}
\begin{figure}[t]
\begin{center}
\includegraphics[width=13cm]{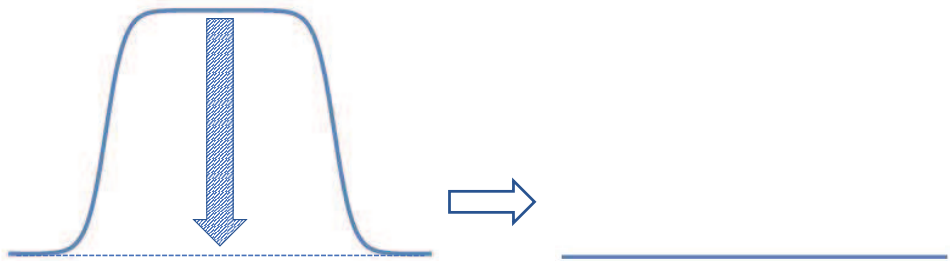} 
\end{center}
\caption{A kink-antikink configuration is unstable, and will decay into the vacuum. }
\label{decay_kink}
\end{figure}

Throughout the paper, we consider a five-dimensional (5D) theory whose fifth dimension is compactified on $S^1$, 
\ie, $y\sim y+L$ ($y$ is the coordinate of $S^1$).
Let us first consider a case of non-gravitational theory. 
Naively, the domain-wall configuration seems to be unstable 
because a kink configuration should be paired with an antikink configuration 
due to the periodic boundary condition, and the kink-antikink pair will decay into the vacuum 
(see Fig.~\ref{decay_kink}). 
This problem can be solved by introducing a topological winding~\cite{Maru:2001gf}. 
As an example, we consider the following model of a real scalar field~$\Phi$. 
\bea
 \cL \eql -\frac{1}{2}\der^M\Phi\der_M\Phi-V(\Phi),  \label{cL:global}
\eea
where $M=0,1,\cdots,4$, and 
\be
 V(\Phi) = C_1-C_2\sin^2\brkt{\frac{\pi\Phi}{v}}, \label{V:global}
\ee
where $C_1$, $C_2$ and $v$ are positive constants. 
The mass dimensions of the parameters are 
\be
 [C_1] = [C_2] = 5, \;\;\;\;\;
 [v] = \frac{3}{2}. 
\ee
The periodic potential~(\ref{V:global}) has the following vacua.  
\be
 \vev{\Phi} = \brkt{n+\frac{1}{2}}v. \;\;\;\;\; (n\in{\mathbb Z})
\ee
Here we assume that the field space is compact, and take the following identification. 
\be
 \Phi \sim \Phi+nv, \;\;\;\;\; (n\in{\mathbb Z})
\ee
which is consistent with the scalar potential~(\ref{V:global}). 
In this case, $n$-domain-wall solution corresponds to the field configuration with a winding number~$n$. 
\be
 \Phi(y+L) = \Phi(y)+nv, \label{bd:winding}
\ee
where $y\equiv x^4$.

\subsection{Domain-wall solution}
\begin{figure}[t]
\begin{center}
\includegraphics[width=8cm]{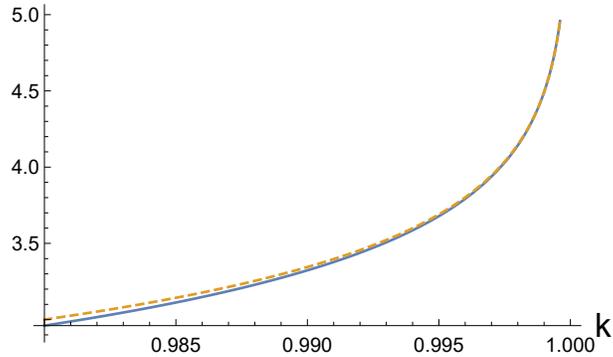} 
\end{center}
\caption{The plots of $kK(k)$ (solid line) and $\ln(4/\sqrt{1-k^2})$ (dashed line).}
\label{kK}
\end{figure}

From (\ref{cL:global}) with (\ref{V:global}), the equation of motion is 
\be
 \der^M\der_M\Phi+\frac{\pi C_2}{v}\sin\brkt{\frac{2\pi\Phi}{v}} = 0. 
\ee
A background solution that is independent of the 4D coordinates~$x^\mu$ is found to be
\be
 \Phi_{\rm bg}(y) = \pm\frac{v}{\pi}{\rm am}\brkt{\frac{\pi\sqrt{2C_2}}{kv}(y-y_0),k}, \label{Phi:global}
\ee
where ${\rm am}(u,k)$ denotes the Jacobi amplitude (see Appendix~\ref{Ja}). 
The integration constants are $k$ and $y_0$. 
In the following, we choose $y_0$ to be zero by shifting the origin of $y$. 
Namely, $\Phi_{\rm bg}(0)=0$. 
Using the property~(\ref{shift:am}), the winding condition~(\ref{bd:winding}) is translated into
\be
 \frac{\pi\sqrt{2C_2}}{kv}\cdot L = 2nK(k),  \label{period}
\ee
where $K(k)$ is the complete elliptic integral of the first kind.  
This is the equation that determines the value of $k$. 
The function~$kK(k)$ is approximated as
\be
 kK(k) \simeq \ln\frac{4}{\sqrt{1-k^2}}, \label{app:kK}
\ee
for $k\simeq 1$ (see Fig.~\ref{kK}). 
Thus, the size of the extra dimension~$L$ is expressed as
\be
 L \simeq \frac{2nv}{\pi\sqrt{2C_2}}\ln\frac{4}{\sqrt{1-k^2}}
 = \frac{nv}{\pi\sqrt{2C_2}}\brkt{s+4\ln 2}, \label{app:L}
\ee
where $k^2\equiv 1-e^{-s}$. 

\begin{figure}[t]
\begin{center}
\includegraphics[width=8cm]{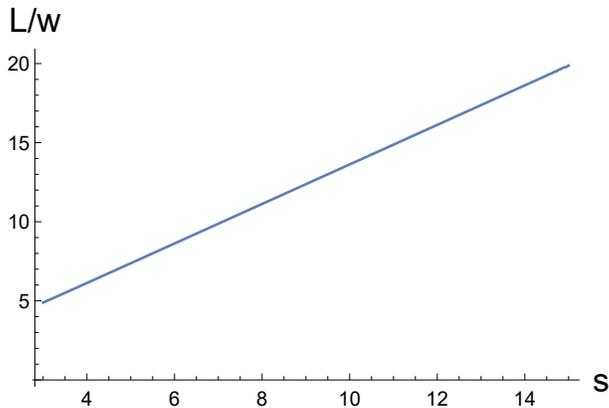} 
\end{center}
\caption{The ratio of the size of the extra dimension~$L$ to the domain-wall width~$w$ 
as a function of $s=\ln\brc{1/(1-k^2)}$ in the case of $n=2$.}
\label{rs}
\end{figure}

When $L$ is large enough (\ie, $k\simeq 1$), the derivative of $\Phi_{\rm bg}(y)$ at the origin is 
determined only by $C_2$. 
\be
 \Phi_{\rm bg}'(0) = \pm\frac{\sqrt{2C_2}}{k} \simeq \pm\sqrt{2C_2}.  
\ee
where the prime denotes the $y$-derivative. 
Thus, the width of the domain wall~$w$ is insensitive to the parameter~$k$, 
while the size of the extra dimension~$L$ depends on it as (\ref{app:L}).  
If we identify the internal region of the domain wall as the range of $y$ where 
\be
 \abs{\Phi_{\rm bg}(y)-\brkt{n+\frac{1}{2}}v}\geq \frac{v}{10}, \;\;\;\;\;
 \brkt{n\in{\mathbb Z}}
\ee
the ratio of $L$ to the width~$w$ is shown in Fig.~\ref{rs} in the two-domain-wall case. 
The horizontal axis denotes $s$ defined below (\ref{app:L}). 

In the limit of $L\to\infty$ (\ie, $k\to 1$), 
the solution~$\Phi_{\rm bg}(y)$ approaches the domain wall in the non-compact extra dimension. 
\be
 \Phi_{\rm bg}(y) \to \pm\frac{v}{\pi}\arctan\brkt{\sinh\brkt{\frac{\pi\sqrt{2C_2}}{v}y}}, 
\ee
which connects the vacua~$\Phi=\pm v/2$. 

As (\ref{Phi:global}) shows, $\Phi_{\rm bg}(y)$ connects the adjacent two vacua, 
and the domain walls are located at equal distances. 
This can be understood from the known fact that each kink configuration feels 
the repulsive force from other kink configurations.

\section{Domain walls in gravitational theory} \label{DWinG}
Now we extend the previous model~(\ref{cL:global}) to the gravitational theory. 
\be
 \cL = \sqrt{-G^{(5)}}\brc{\frac{1}{2\kp_5^2}\cR^{(5)}-\frac{1}{2}g^{MN}\der_M\Phi\der_N\Phi-V(\Phi)}, 
\ee
where $\kp_5$ is the 5D gravitational coupling, $G^{(5)}$ is the determinant of the 5D metric~$g_{MN}$, 
and $\cR^{(5)}$ is the 5D Ricci scalar.  
The  equations of motion are given by
\bea
 \cR_{MN}^{(5)}-\frac{1}{2}g_{MN}\cR^{(5)} \eql \kp_5^2\sbk{\der_M\Phi\der_N\Phi
 +g_{MN}\brc{-\frac{1}{2}g^{PQ}\der_P\Phi\der_Q\Phi-V(\Phi)}}, \nonumber\\
 g^{MN}\nabla_M\der_N\Phi-\frac{dV}{d\Phi}(\Phi) \eql 0.  \label{EOM}
\eea

As we will show below, there is no static domain-wall solution, 
in contrast to the non-gravitational theory.

\subsection{Absence of static domain-wall solutions} \label{DWinG:1}
In order to search for a static domain-wall solution, we take the following ansatz 
for the metric and the field. 
\bea
 ds^2 \eql e^{2\sgm(y)}\eta_{\mu\nu}dx^\mu dx^\nu+dy^2, \nonumber\\
 \Phi \eql \Phi(y).  \label{bgd:static}
\eea
Then, the equations in (\ref{EOM}) becomes
\bea
 &&3\sgm''+6\sgm^{\prime 2} = \kp_5^2\brc{-\frac{1}{2}\Phi^{\prime 2}-V(\Phi)}, \nonumber\\
 &&6\sgm^{\prime 2} = \kp_5^2\brc{\frac{1}{2}\Phi^{\prime 2}-V(\Phi)}, \nonumber\\
 &&\Phi''+4\sgm'\Phi'-\frac{dV}{d\Phi}(\Phi) = 0, \label{EOM:static}
\eea
which come from the $(\mu,\nu)$- and $(y,y)$-components of the 5D Einstein equation, 
and the field equation for $\Phi$, respectively. 
The prime denotes the $y$-derivative. 
The first two equations can be rewritten as
\bea
 3\sgm'' \eql -\kp_5^2\Phi^{\prime 2}, \nonumber\\
 \sgm^{\prime 2} \eql \frac{\kp_5^2}{6}\brc{\frac{1}{2}\Phi^{\prime 2}-V(\Phi)}. 
 \label{EOM:redef}
\eea

When the widths of the domain walls are small enough compared to the size of the extra dimension, 
each wall can be regarded as a 3-brane with a positive tension. 
It is well-known that a periodic warp function~$\sgm(y)$ cannot be obtained 
by introducing only positive-tension branes~\cite{Oda:1999di,Hatanaka:1999ac}. 
A negative-tension brane is necessary for a static multi-brane solution in the compact extra dimension. 
However, such a brane cannot be obtained from any kinds of domain walls. 
Therefore, we expect that there is no static domain-wall solution in our setup. 
We can show that this is indeed the case as follows.
  
For our purpose, it is convenient to use the first-order formalism~\cite{Skenderis:1999mm,DeWolfe:1999cp,Afonso:2006gi}. 
We introduce the function~$W=W(\Phi)$, which satisfies 
\be
 \sgm'(y) = -\frac{\kp_5^2}{3}W(\Phi), \;\;\;\;\;
 \Phi'(y) = W_\Phi(\Phi),  \label{suppot}
\ee
where $W_\Phi\equiv dW/d\Phi$. 
These are consistent with the first equation in (\ref{EOM:redef}). 
From the second equation in (\ref{EOM:redef}), the potential is expressed as
\bea
 V(\Phi) \eql \frac{1}{2}\Phi^{\prime 2}-\frac{6}{\kp_5^2}\sgm^{\prime 2} \nonumber\\
 \eql \frac{1}{2}W_\Phi^2(\Phi)-\frac{2\kp_5^2}{3}W^2(\Phi).  \label{V:W}
\eea
By using this expression, 
we can show that (\ref{suppot}) is also consistent with the last equation in (\ref{EOM:static}). 
If we choose $W(\Phi)$ as
\be
 W(\Phi) = \brkt{\frac{\pi^2}{2v^2}+\frac{2\kp_5^2}{3}}^{-1/2}\sqrt{C_2}\sin\brkt{\frac{\pi\Phi}{v}}, 
\ee
we can reproduce the potential~(\ref{V:global}), up to a constant term. 

From the second equation in (\ref{suppot}), we have~\footnote{
We assume that $\Phi(y)$ is a monotonic function of $y$. 
} 
\be
 \int_0^\Phi\frac{d\tl{\Phi}}{W_\Phi(\tl{\Phi})} = y. 
\ee
This indicates that $W_\Phi(\Phi)$ is a periodic function with the period~$v$.\footnote{ 
$\Phi(y)$ is unbounded due to the boundary condition~(\ref{bd:winding}). 
}
Besides, from the expression~(\ref{V:W}), $\abs{W(\Phi)}$ must be bounded from above. 
Therefore, $W(\Phi)$ must satisfy the following conditions. 
\begin{itemize}
\item $W(\Phi)$ is a periodic function of $\Phi$ with the period $v$.  

\item The following relation must be held. 
\be
 \int_0^v\frac{d\tl{\Phi}}{W_\Phi(\tl{\Phi})} = L < \infty.  \label{int:Wp}
\ee
\end{itemize}
From the first requirement, it follows that
\be
 \int_0^v d\tl{\Phi}\;W_\Phi(\tl{\Phi}) = W(v)-W(0) = 0. 
\ee
Thus, $W_\Phi(\Phi)$ has a simple zero in $[0,v]$. 
Hence, the integral in the LHS of (\ref{int:Wp}) diverges, and the requirement~(\ref{int:Wp}) cannot be satisfied. 
Namely, there is no static solution that has a non-zero winding number.

\subsection{Non-static domain-wall solutions} \label{DWinG:2}
As we showed, any domain walls must depend on time. 
In order to see this time-evolution, 
we take the following metric ansatz~\cite{Takamizu:2006gm,Omotani:2011un}. 
\be
 ds^2 = e^{2A(t,z)}\brkt{-dt^2+dz^2}+e^{2B(t,z)}d\vec{x}^2.  \label{5Dmetric}
\ee
Then, (\ref{EOM}) is translated into~\footnote{
Here, $\Phi$ is canonically normalized, in contrast to Ref.~\cite{Takamizu:2006gm}. 
} 
\bea
 \ddot{A} \eql A''+3\dot{B}^2-3B^{\prime 2}
 -\frac{\kp_5^2}{2}\brc{\dot{\Phi}^2-\Phi^{\prime 2}+\frac{2}{3}e^{2A}V(\Phi)}, 
 \nonumber\\
 \ddot{B} \eql B''-3\dot{B}^2+3B^{\prime 2}+\frac{2\kp_5^2}{3}e^{2A}V(\Phi), \nonumber\\
 \ddot{\Phi} \eql \Phi''-3\dot{B}\dot{\Phi}+3B'\Phi'-e^{2A}\frac{dV}{d\Phi}(\Phi),  \label{t-EOM:1}
\eea
with 
\bea
 \dot{B}B'-A'\dot{B}-\dot{A}B'+\dot{B}' \eql -\frac{\kp_5^2}{3}\dot{\Phi}\Phi', \nonumber\\
 2B^{\prime 2}+B''-A'B'-\dot{A}\dot{B}-\dot{B}^2
 \eql -\frac{\kp_5^2}{6}\brc{\dot{\Phi}^2+\Phi^{\prime 2}+2e^{2A}V(\Phi)}, \label{t-EOM:2}
\eea
where the dot and the prime denote the derivatives with respect to $t$ and $z$, respectively. 
If the $(M,N)$-component of the 5D Einstein equation is denoted as $\cG_{MN}=0$, 
the first two equations in (\ref{t-EOM:1}) are $\cG_{ii}+\frac{2}{3}(\cG_{tt}-\cG_{zz})=0$ 
and $\frac{1}{3}(\cG_{tt}-\cG_{zz})=0$, while the equations in (\ref{t-EOM:2}) 
are $-\frac{1}{3}\cG_{tz}=0$ and $-\frac{1}{3}\cG_{tt}=0$, respectively.  
The other components do not provide non-trivial equations. 
Note that the equations in (\ref{t-EOM:2}) do not contain the second-order derivative with respect to time. 
Thus, they are treated as the constraints in the numerical calculation for the time evolution. 

In the limit of decompactifying the extra dimension, the static solution is allowed. 
In fact, neglecting the time-dependences of the background functions,\footnote{
The two functions~$A(t,z)$ and $B(t,z)$ should be reduced to the same function~$\tl{\sgm}(z)$ 
if we require the 4D Lorentz invariance.  
} 
\bea
 A(t,z), \; B(t,z) \toa \tl{\sgm}(z), \nonumber\\
 \Phi(t,z) \toa \tl{\Phi}(z),  \label{static_limit}
\eea
and redefining the extra-dimensional coordinate as
\be
 y \equiv \int_0^z d\tl{z}\;e^{\tl{\sgm}(\tl{z})},  \label{def:y}
\ee
we reproduces the metric~(\ref{bgd:static}) and the equations of motion~(\ref{EOM:static}) 
after rewriting $\sgm(y)\equiv \tl{\sgm}(z)$ and $\Phi(y)\equiv \tl{\Phi}(z)$.\footnote{
Note that $\der_z = e^{\sgm(y)}\der_y$. 
} 
(See Appendix~\ref{static_limit_EOM}.) 

The physical size of the extra dimension at time~$t$ is 
\be
 \tl{L}_{\rm phys}(t) = \int_0^L dz\;e^{A(t,z)}.  \label{tlL_phys}
\ee
For larger length scales than $\tl{L}_{\rm phys}(t)$, the spacetime becomes 4D-like. 
From (\ref{5Dmetric}), the line elements along the time and the 3D space directions 
that are measured on the wall are 
\bea
 (\Delta s_t)_{\rm 4D} \eql \int_0^L dz\;\abs{f_{\rm ob}(t,z)}^2e^{A(t,z)}\Delta t
 \equiv e^{\bar{A}(t)}\Delta t, \nonumber\\
 (\Delta s_{\vec{x}})_{\rm 4D} \eql \int_0^L dz\;\abs{f_{\rm ob}(t,z)}^2e^{B(t,z)}\Delta\vec{x}
 \equiv e^{\bar{B}(t)}\Delta\vec{x},  \label{def:line_el}
\eea
where $f_{\rm ob}(t,z)$ denotes the wave function of the observer in the extra dimension at time~$t$. 
Thus, the effective 4D metric is 
\be
 ds_{\rm 4D}^2 = -e^{2\bar{A}(t)}dt^2+e^{2\bar{B}(t)}d\vec{x}^2. 
\ee

The cosmic time~$\tau$ is thus given by
\be
 \tau \equiv \int_0^t d\tl{t}\;e^{\bar{A}(\tl{t})}.  \label{def:tau}
\ee
In terms of $\tau$, the 4D metric is rewritten as
\be
 ds_4^2 = -d\tau^2+a^2(\tau)d\vec{x}^2, 
\ee
where the scale factor~$a(\tau)$ is defined by
\be
 a(\tau) \equiv \exp\brc{\bar{B}(t(\tau))}.  \label{def:a}
\ee
Here, $t(\tau)$ is the inverse function of (\ref{def:tau}). 
The Hubble parameter~$H$ is then expressed as
\be
 H(\tau) \equiv \frac{\der_\tau a(\tau)}{a(\tau)} 
 = e^{-\bar{A}(t)-\bar{B}(t)}\der_t e^{\bar{B}(t)}. 
\ee
In terms of $\tau$, the size of the extra dimension~(\ref{tlL_phys}) is rewritten as
\be
 L_{\rm phys}(\tau) \equiv \tl{L}_{\rm phys}(t(\tau)). 
\ee

\section{Numerical results} \label{num_result}
In this section, we show our numerical results. 
We focus on the case of two domain walls, as an example. 
As an initial configuration, we choose the static solution in the non-gravitational case. 
Namely, $A=B=0$ and $\Phi=\Phi_{\rm bg}$, which is shown in (\ref{Phi:global}) 
with (\ref{period}) for $n=2$. 
Then, since we work in the gravitational theory, 
the $t$-derivatives of the fields must have nontrivial profiles 
due to the constraints in (\ref{t-EOM:2}). 
Specifically, the initial configuration is given by 
\bea
 A(t=0,z) \eql B(t=0,z) = 0, \nonumber\\
 \Phi(t=0,z) \eql \Phi_0(z) \equiv \frac{v}{\pi}{\rm am}\brkt{\frac{4K(k)}{L}z,k}, \nonumber\\
 \dot{A}(t=0,z) \eql \frac{\kp_5^2}{6b_0}\brc{\Phi_0^{\prime 2}+2V(\Phi_0)}-b_0 \nonumber\\
 \eql \frac{\kp_5^2}{3b_0}\brc{\frac{C_2}{k^2}-C_1+2V(\Phi_0)}-b_0, \nonumber\\
 \dot{B}(t=0,z) \eql b_0, \;\;\;\;\;
 \dot{\Phi}(t=0,z) = 0,  \label{ini_conf}
\eea
where $b_0$ is a real constant, and the constant~$k$ is determined by the model parameters through 
\be
 \frac{\pi L\sqrt{2C_2}}{v} = 4kK(k). 
\ee
At the second equality for $\dot{A}(0,z)$, we have used that
\bea
 \Phi_0'(z) \eql \frac{v}{\pi}\cdot\frac{\pi\sqrt{2C_2}}{kv}{\rm dn}\brkt{\frac{\pi\sqrt{2C_2}}{kv}z,k} \nonumber\\
 \eql \frac{\sqrt{2C_2}}{k}\brc{1-k^2\sin^2\brkt{{\rm am}\brkt{\frac{\pi\sqrt{2C_2}}{kv}z,k}}}^{1/2} \nonumber\\
 \eql \frac{\sqrt{2C_2}}{k}\brc{1-\frac{k^2}{C_2}\brkt{C_1-V(\Phi_0)}}^{1/2}. 
\eea
(See (\ref{der_am}), (\ref{def:sns}) and (\ref{V:global}).)

In order to calculate the effective 4D metric, we choose the wave function of the observer as
\be
 f_{\rm ob}(t,z) = N_{\rm ob}(t)\Phi'(t,z), 
\ee
where $N_{\rm ob}(t)$ is the normalization factor that satisfies 
\be
 \int_0^L dz\;f_{\rm ob}^2(t,z) = 1. 
\ee

In the following, the parameters in the model and the initial configuration are chosen as
\be
 \kp_5=1.0, \;\;\;\;\;
 C_2 = 0.1, \;\;\;\;\;
 L = 20, \;\;\;\;\;
 k = 1-10^{-8}, \;\;\;\;\;
 b_0 = 1.0, 
 \label{prm_choice}
\ee
which lead to $v=0.685$. 
As we will show below, the time evolution of the configuration depends on 
the minimal value of the potential, 
\be
 V_{\rm min} \equiv C_1-C_2. 
\ee

\subsection{Case of $\bdm{V_{\rm min}\geq 0}$}
First we consider a case of $V_{\rm min}\geq 0$ ($C_1\geq 0.1$). 
Since the qualitative behavior of the background configuration does not depend 
much on a specific value of $V_{\rm min}$ as long as $V_{\rm min}\geq 0$, 
we mainly focus on the case of $V_{\rm min}=0$ in this subsection. 

\begin{figure}[H]
\begin{center}
\includegraphics[width=60mm]{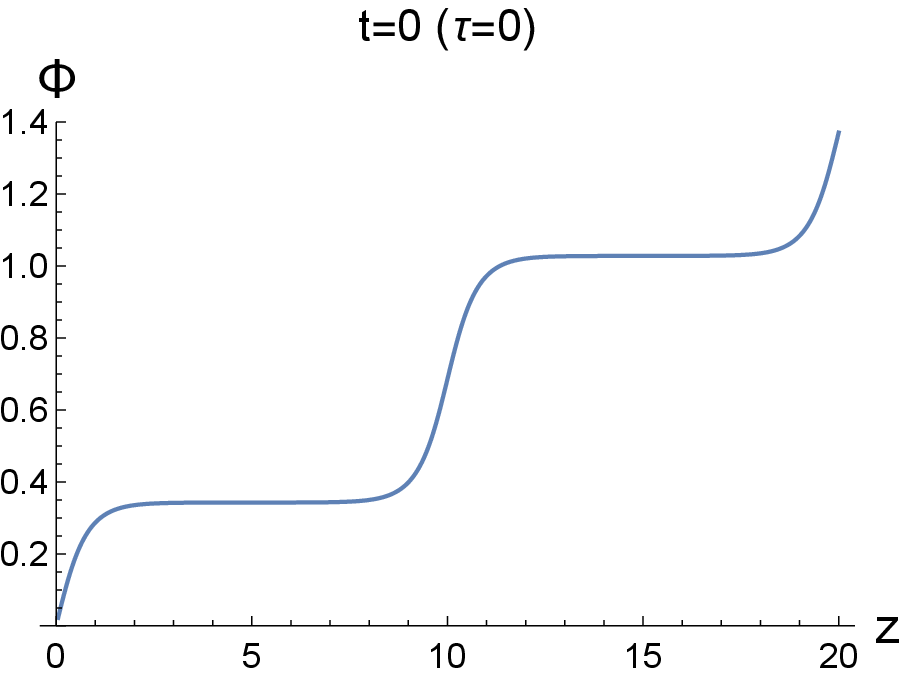} \hspace{8mm}
\includegraphics[width=60mm]{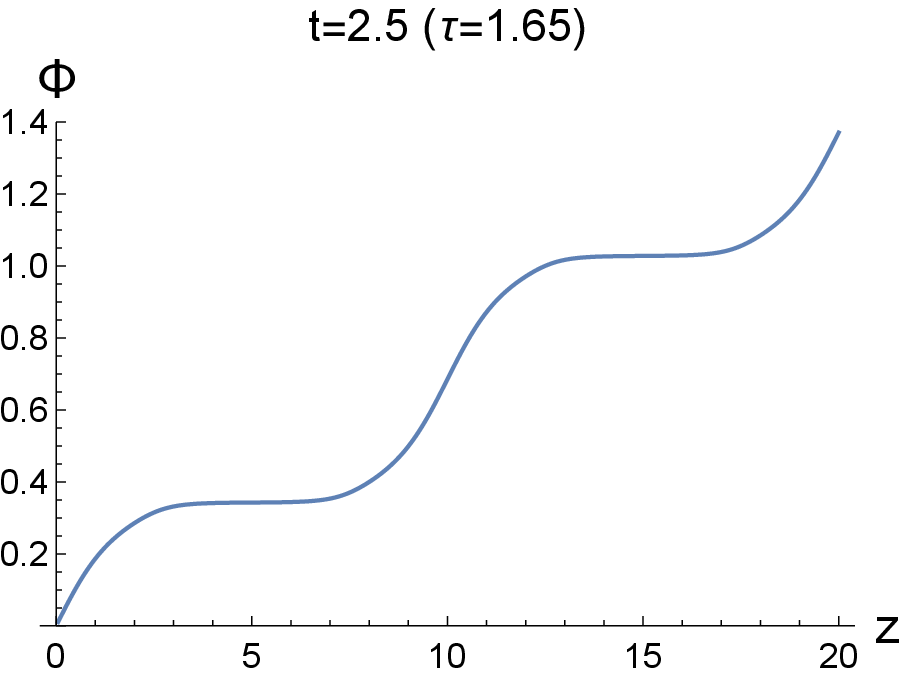} \\
\vspace{5mm}
\includegraphics[width=60mm]{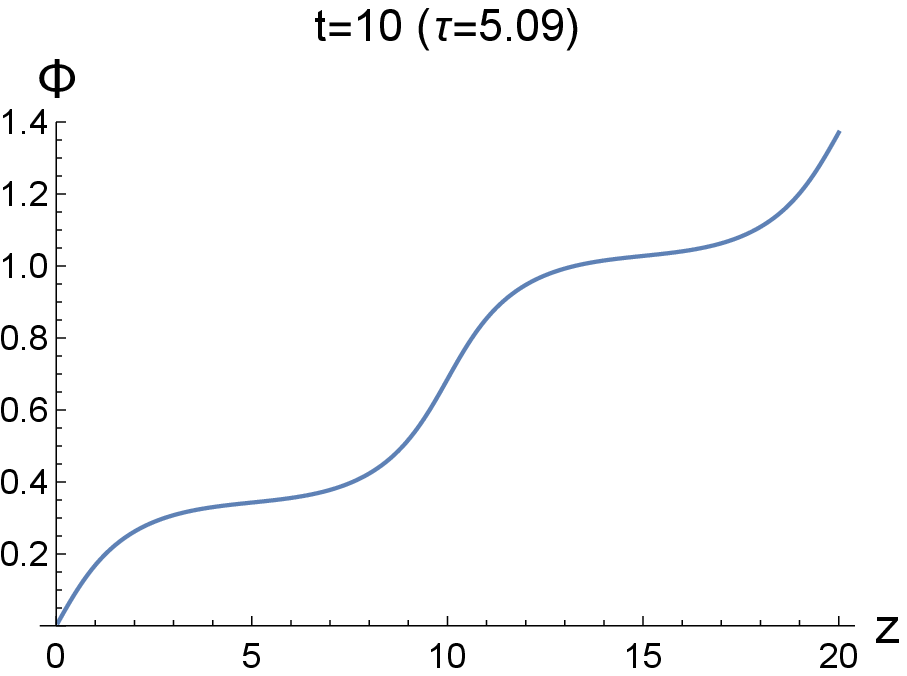} \hspace{8mm}
\includegraphics[width=60mm]{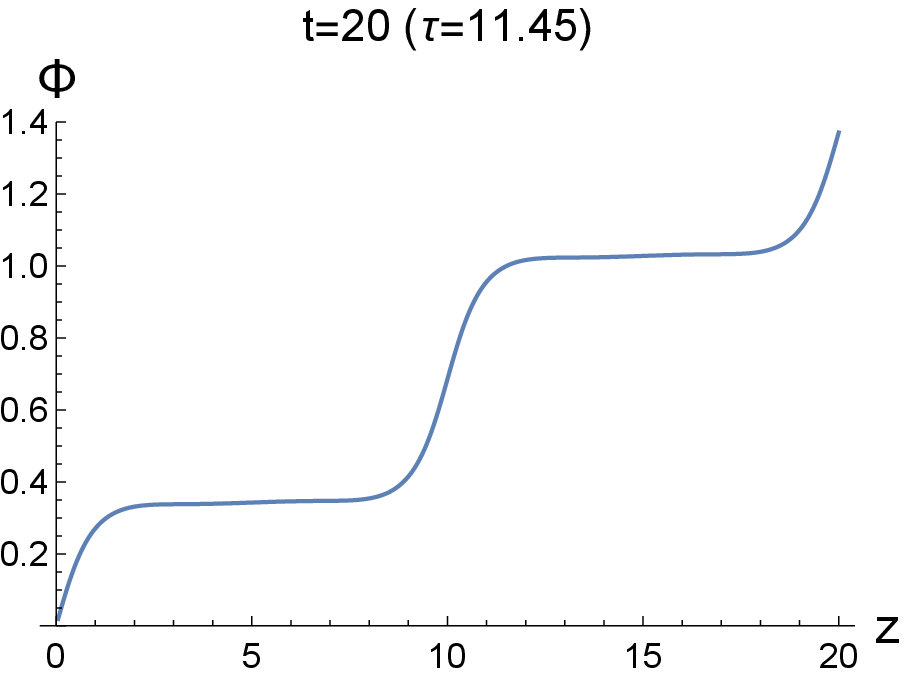} \\
\vspace{5mm}
\includegraphics[width=60mm]{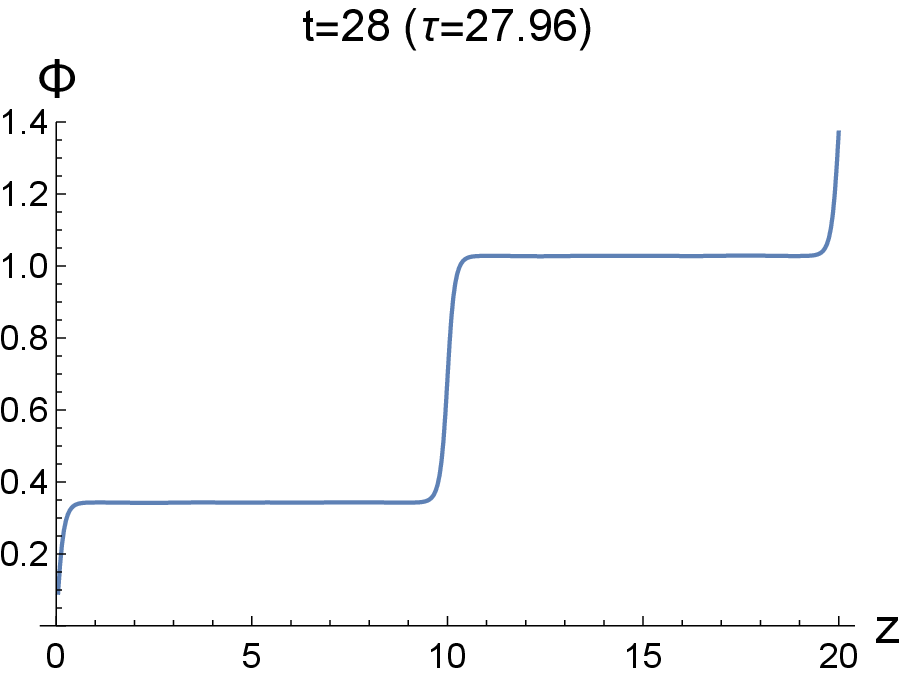} \hspace{8mm}
\includegraphics[width=60mm]{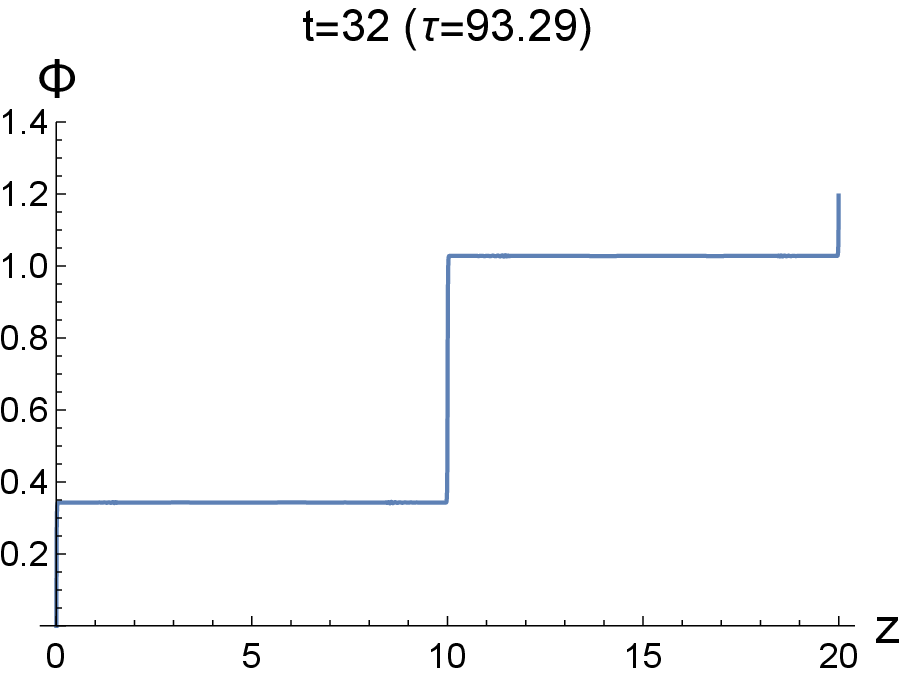}
\end{center}
\caption{The profiles of $\Phi$ at various times. 
The parameters are chosen as (\ref{prm_choice}) and $V_{\rm min}=0$.  }
\label{profile:Phi}
\end{figure}

\begin{figure}[H]
\begin{center}
\includegraphics[width=80mm]{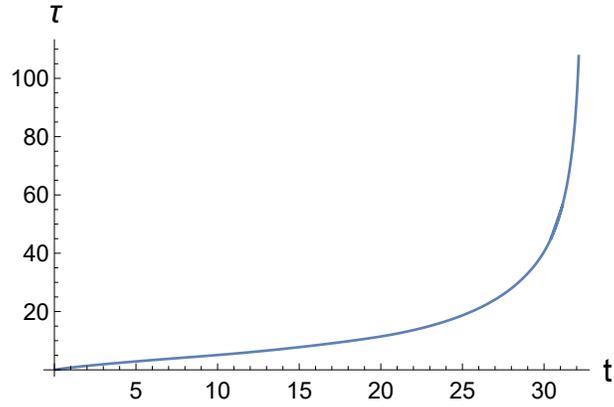}
\end{center}
\caption{The relation between $t$ and $\tau$. 
The parameters are chosen as (\ref{prm_choice}) and $V_{\rm min}=0$.  }
\label{rel:t-tau}
\end{figure}

Fig.~\ref{profile:Phi} shows the profile of $\Phi$ at various times. 
We can see that the scalar configuration once loses the kink shape and approaches a linear function of $z$. 
Then, after some time, it starts to form the kink configuration again, and approaches the (periodic) step 
function as $\tau\to \infty$. 
From Fig.~\ref{profile:Phi}, it seems that the scalar configuration reaches the singular step-function 
profile at a finite time~$t\simeq 32$. 
However, Fig.~\ref{rel:t-tau} indicates that 
it takes infinite time for the cosmic time~$\tau$. 
Thus, the step-function profile is just an asymptotic configuration. 
Here note that a distance measured by the coordinate~$z$ is not the physical one. 
It should be measured by the proper length. 
As we will see below, the above-mentioned behavior is a result of the expansion of the extra dimension, 
and the physical width of the domain wall itself remains constant. 

\begin{figure}[H]
\begin{center}
\includegraphics[width=70mm]{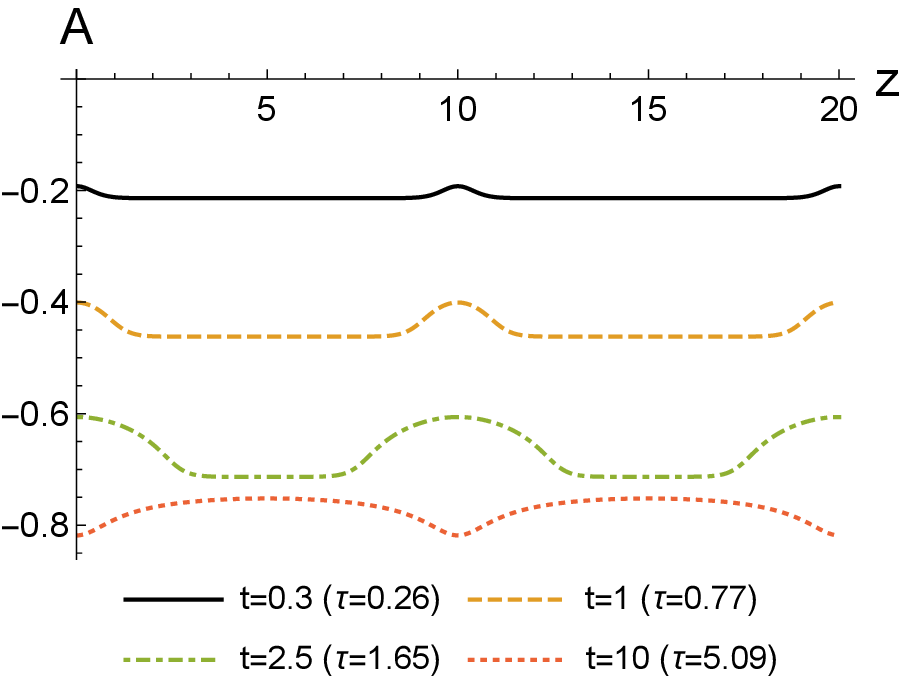} \hspace{8mm}
\includegraphics[width=70mm]{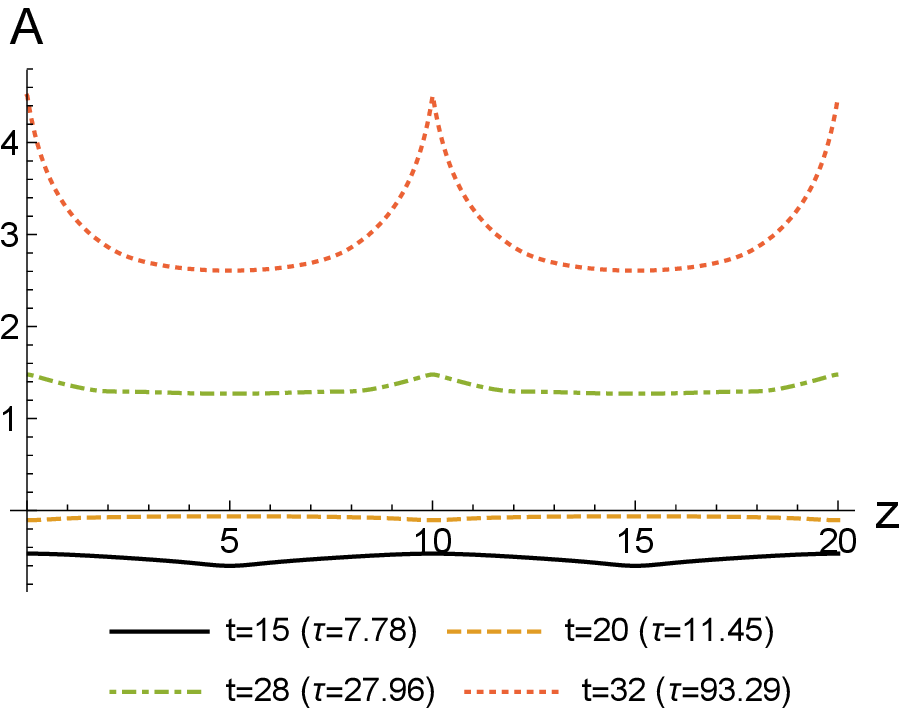} \\
\includegraphics[width=70mm]{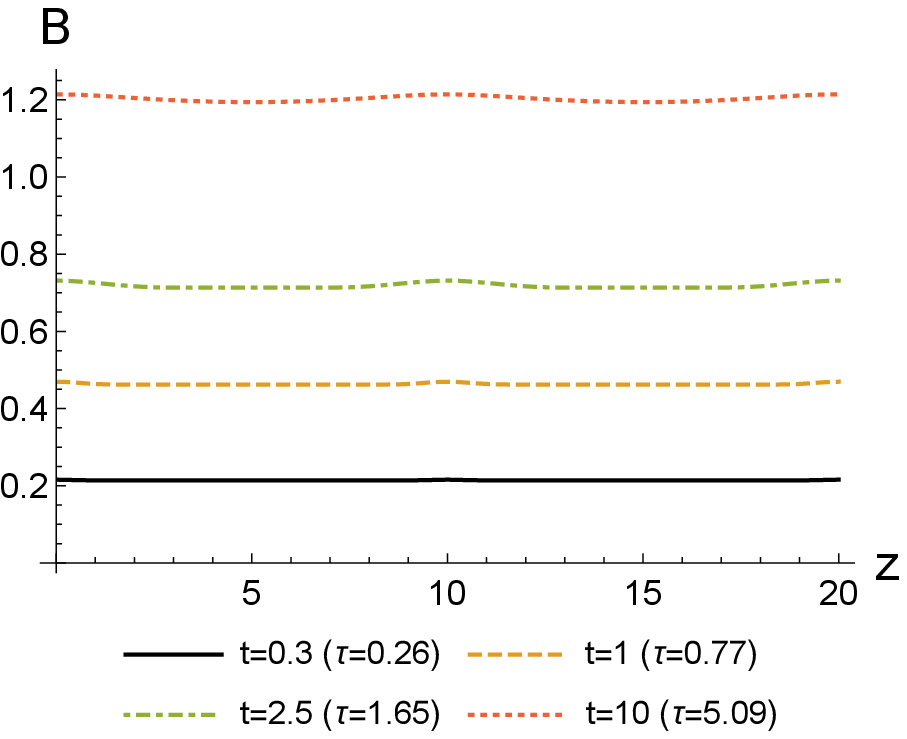} \hspace{8mm}
\includegraphics[width=70mm]{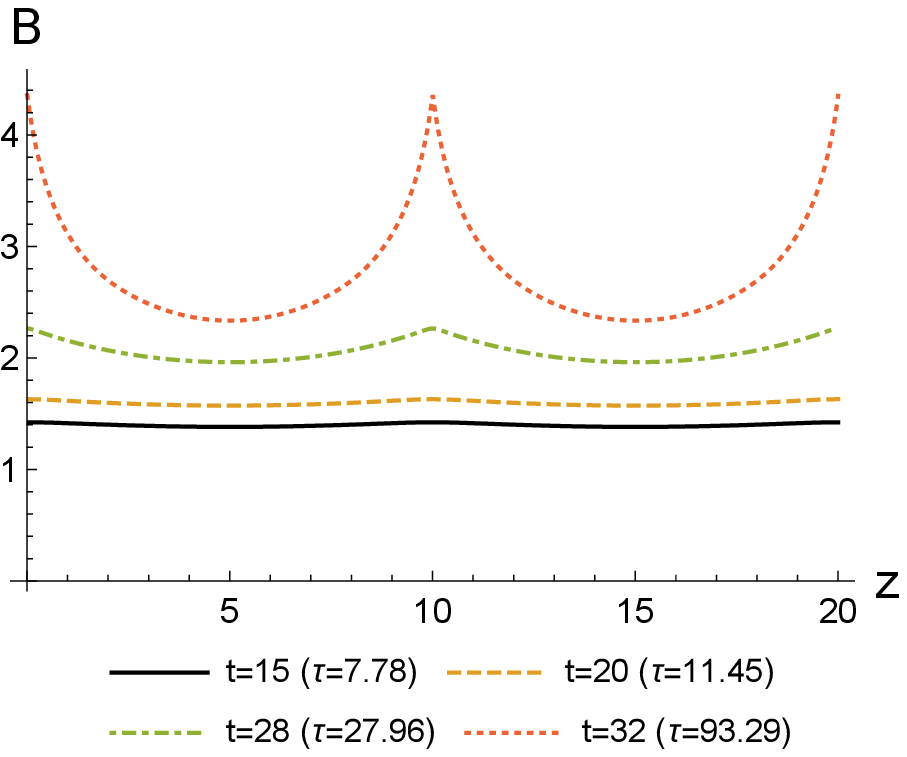} 
\end{center}
\caption{The profiles of $A$ and $B$ at various times.  
The parameters are chosen as (\ref{prm_choice}) and $V_{\rm min}=0$.  }
\label{profile:A}
\end{figure}

Fig.~\ref{profile:A} shows the profiles of $A$ and $B$ at each time. 
The warp factor~$A$ first decreases and stays around $A\simeq -1$ for some time. 
Then, it turns to increase and its $z$-dependence grows. 
At late times, the warp factor~$A$ has peaks at the wall positions. 
This is similar to the Randall-Sundrum model~\cite{Randall:1999vf}. 
The positive tensions of the walls warp the ambient geometry. 
In contrast, the ``3D scale factor''~$B$ monotonically increases with time. 
Its $z$-dependence also grows, and it has peaks at the wall positions, 
which is similar to the behavior of $A$ at late times. 

\begin{figure}[H]
\begin{center}
\includegraphics[width=70mm]{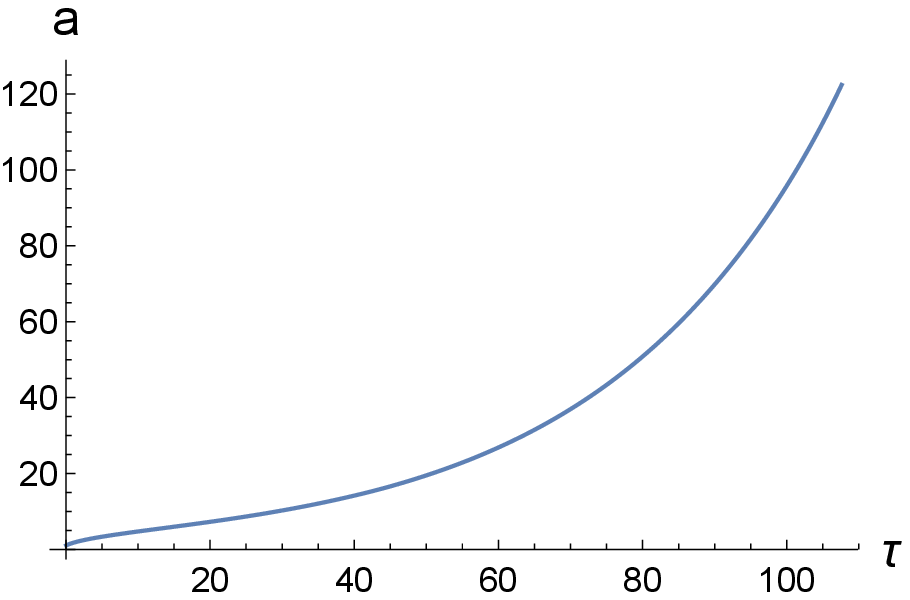} \hspace{8mm}
\includegraphics[width=70mm]{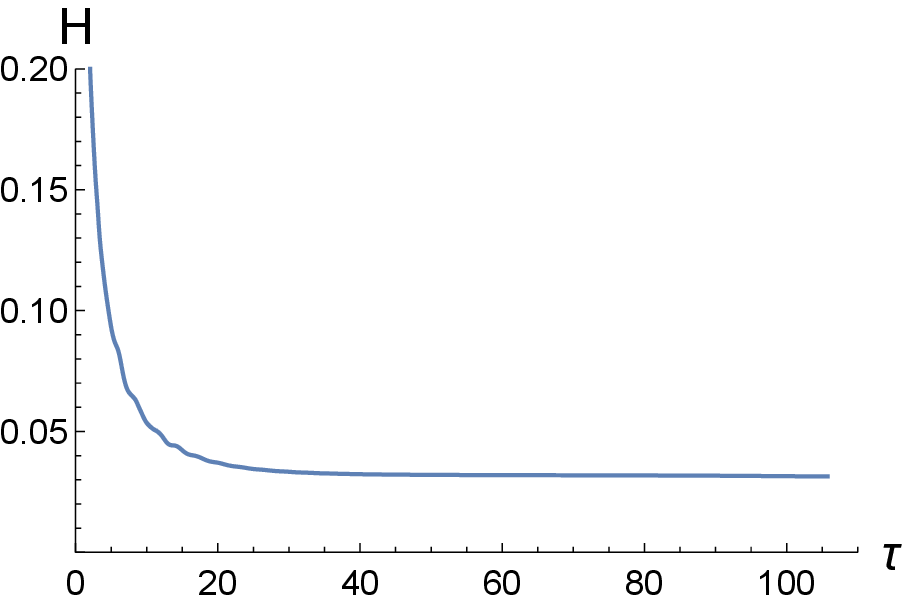} \\
\vspace{5mm}
\includegraphics[width=70mm]{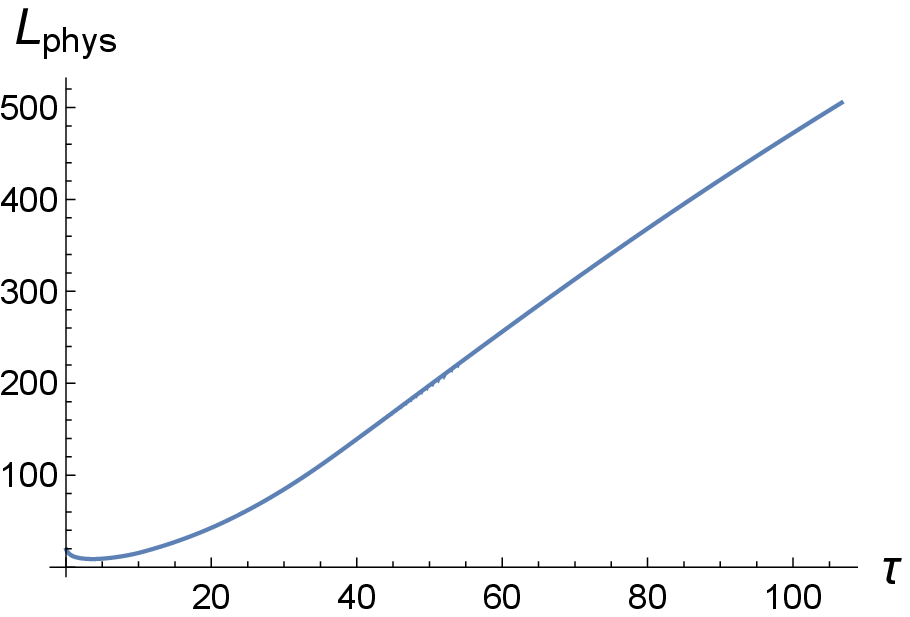} \hspace{8mm}
\includegraphics[width=70mm]{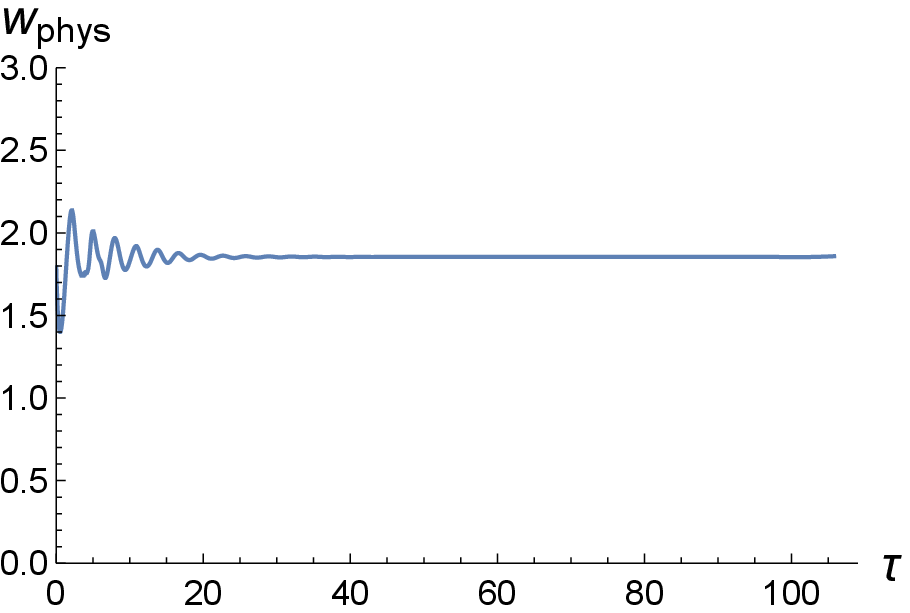}
\end{center}
\caption{The scale factor~$a$, the Hubble parameter~$H$, 
the physical size of the extra dimension~$L_{\rm phys}$, 
and the physical width of the domain wall~$w_{\rm phys}$ as functions of $\tau$. 
The parameters are chosen as (\ref{prm_choice}) and $V_{\rm min}=0$.  }
\label{profile:aH10}
\end{figure}
Fig.~\ref{profile:aH10} shows the scale factor~$a$, the Hubble parameter~$H$, 
the physical size of the extra dimension~$L_{\rm phys}$, 
and the physical width of the domain wall~$w_{\rm phys}$ as functions of $\tau$. 
The Hubble parameter~$H$ decreases rapidly, 
and approaches a positive small value. 
This positive value depends on the constant~$C_2$.\footnote{
For a smaller value of $C_2$, we have a smaller asymptotic value of $H$.  
}
Thus, the accelerated expansion of the non-compact 3D space is realized at late times.  
The size of the extra dimension first shrinks a little, 
and then expands as a linear function of $\tau$. 

As mentioned above, the width of the domain wall must be measured by the proper length. 
Here we identify the domain wall region as the range of $z$ in which 
\be
 \abs{\Phi(t,z)-\brkt{n+\frac{1}{2}}v} \geq \frac{v}{10}. \;\;\;\;\;
 \brkt{n\in{\mathbb Z}}
\ee
Then, the physical wall width~$w_{\rm phys}$ is defined by
\be
 w_{\rm phys}(\tau) \equiv \int_{-z_w}^{z_w} dz\;e^{A(t(\tau),z)}  
 = 2\int_0^{z_w}dz\;e^{A(t(\tau),z)}, 
\ee
where $z_w$ is determined by
\be
 \Phi(z_w) = \frac{4v}{10}. 
\ee
From Fig.~\ref{profile:aH10}, we can see that the width almost remains constant during the evolution. 
Therefore, the behavior of $\Phi$ that approaches the singular step function shown in Fig.~\ref{profile:Phi} 
is understood as a result of the expansion of the extra dimension. 
Namely, although $w_{\rm phys}$ itself does not decrease, the ratio $w_{\rm phys}/L_{\rm phys}$ 
approaches zero because of the linear increase of $L_{\rm phys}$. 

\begin{figure}[H]
\begin{center}
\includegraphics[width=80mm]{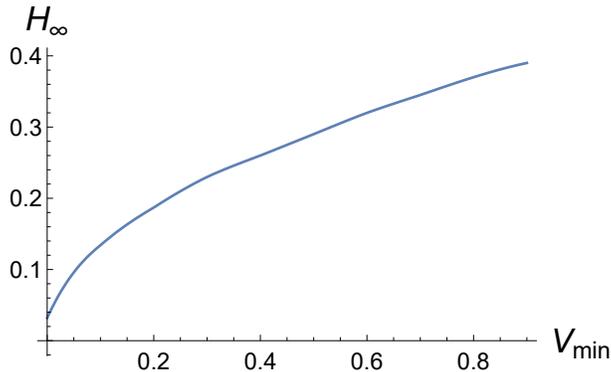}
\end{center}
\caption{The asymptotic value~$H_\infty$ as a function of the minimal value of the potential~$V_{\rm min}=C_1-C_2$. 
The parameters are chosen as (\ref{prm_choice}).  }
\label{Hinf}
\end{figure}
As mentioned above, for positive values of $V_{\rm min}$, the behaviors of $\Phi$, $A$ and $B$ 
are similar to those shown in Figs.~\ref{profile:Phi} and \ref{profile:A}.  
The Hubble parameter~$H$ monotonically decreases, and asymptotically approaches 
a positive constant~$H_\infty$. 
This asymptotic value depends on $V_{\rm min}$. 
Fig.~\ref{Hinf} shows the relation between $H_\infty$ and $V_{\rm min}$.

\subsection{Case of $\bdm{V_{\rm min}<0}$} \label{negV_0}
\begin{figure}[H]
\begin{center}
\includegraphics[width=60mm]{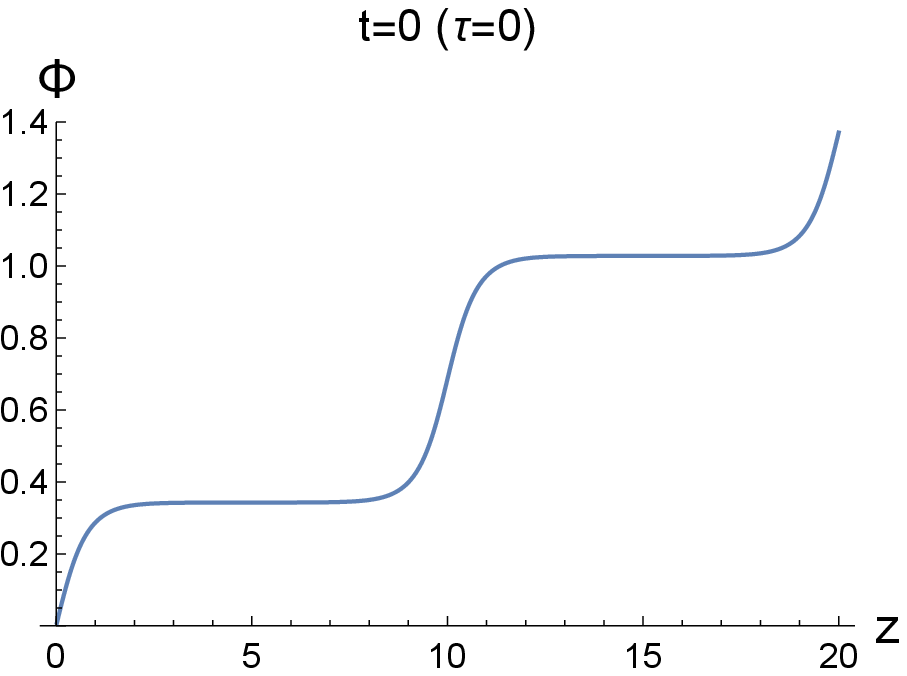} \hspace{8mm}
\includegraphics[width=60mm]{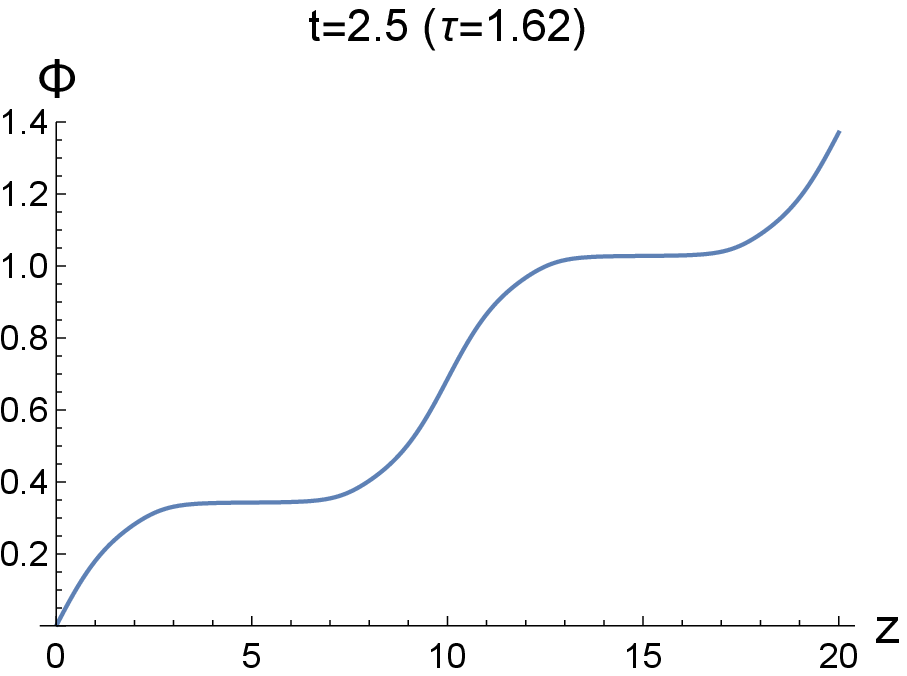} \\
\vspace{5mm}
\includegraphics[width=60mm]{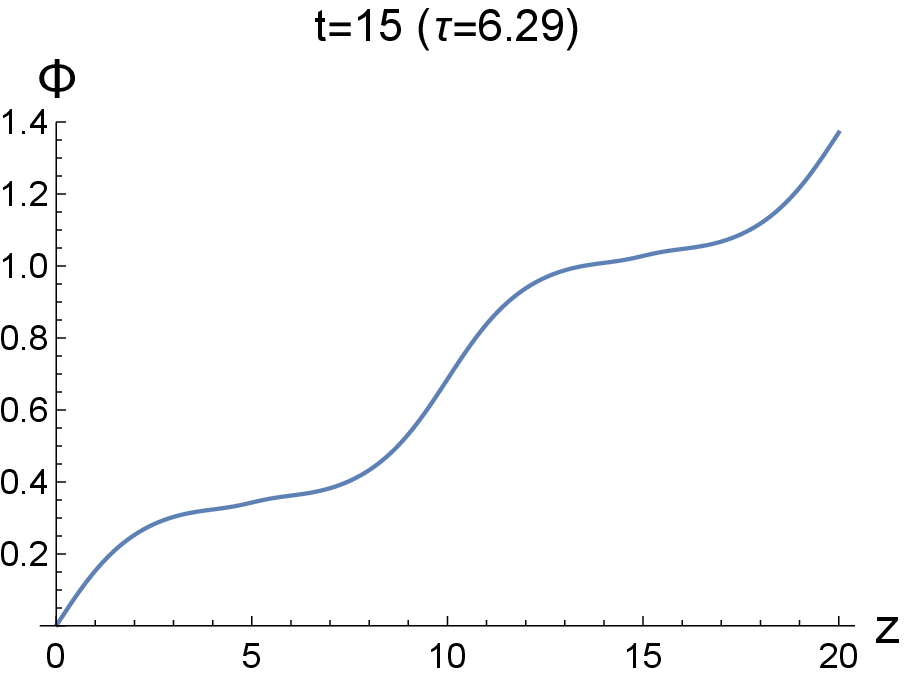} \hspace{8mm}
\includegraphics[width=60mm]{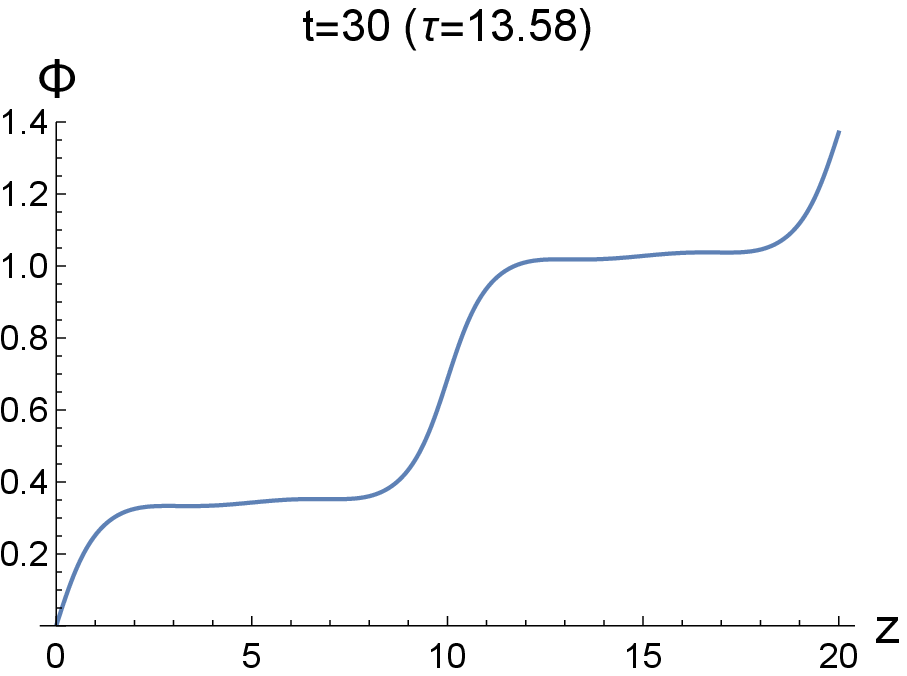} \\
\vspace{5mm}
\includegraphics[width=60mm]{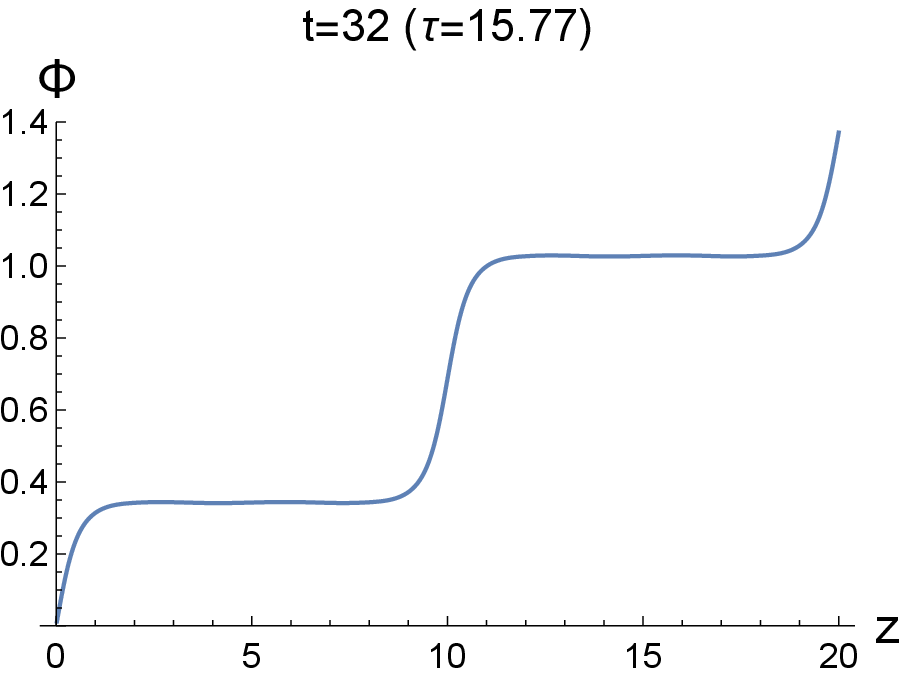} \hspace{8mm}
\includegraphics[width=60mm]{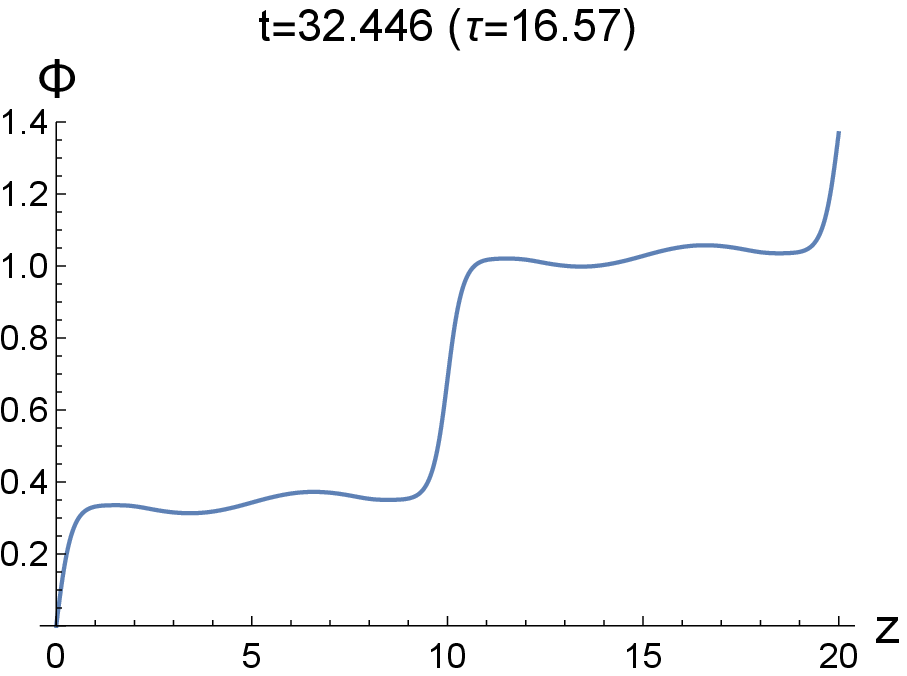}
\end{center}
\caption{The profiles of $\Phi$ at various times. 
The parameters are chosen as (\ref{prm_choice}) and $V_{\rm min}=-0.05$.  }
\label{profile:Phi05}
\end{figure}

\begin{figure}[H]
\begin{center}
\includegraphics[width=70mm]{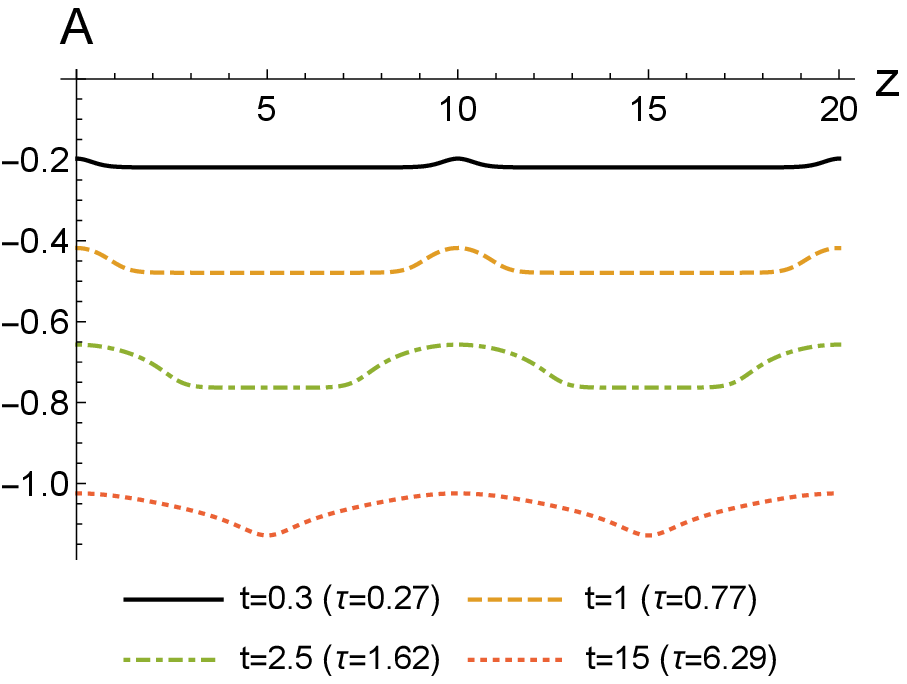} \hspace{8mm}
\includegraphics[width=70mm]{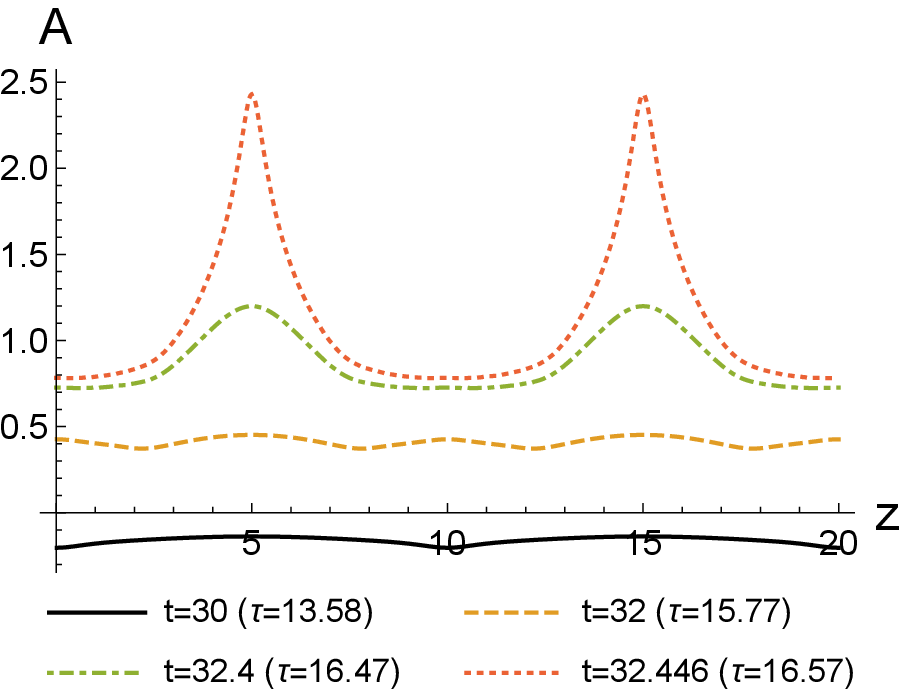} \\
\includegraphics[width=70mm]{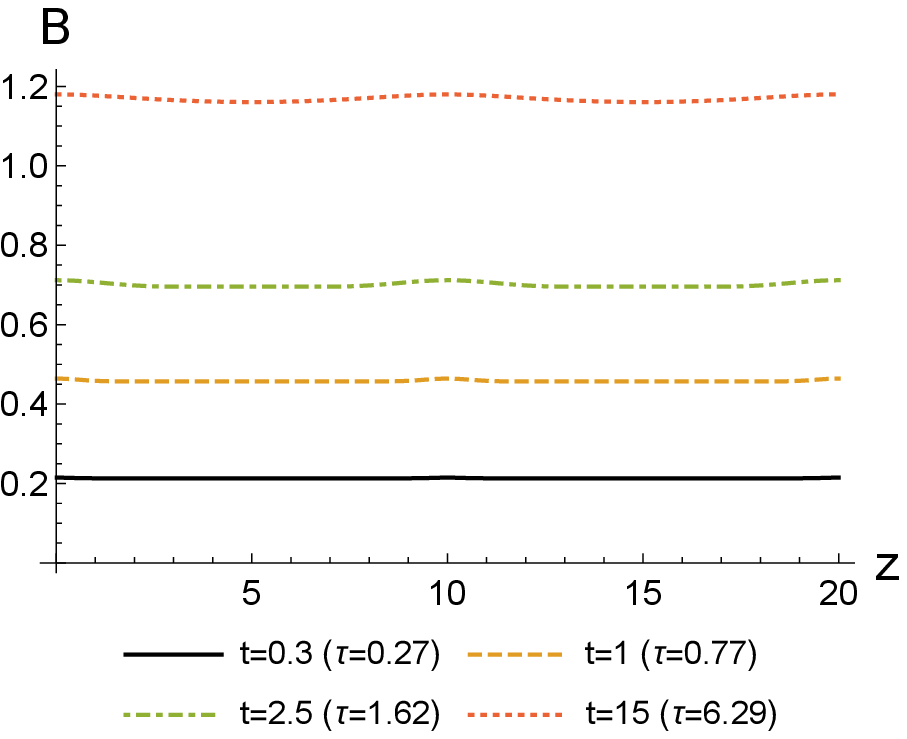} \hspace{8mm}
\includegraphics[width=70mm]{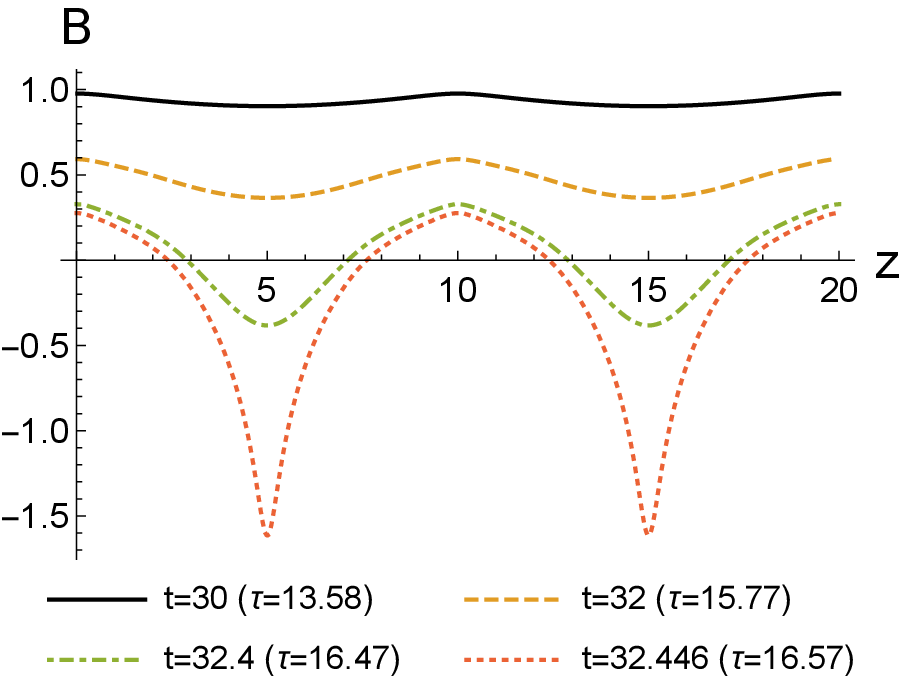} 
\end{center}
\caption{The profiles of $A$ and $B$ at various times.  
The parameters are chosen as (\ref{prm_choice}) and $V_{\rm min}=-0.05$.  }
\label{profile:A05}
\end{figure}

\begin{figure}[H]
\begin{center}
\includegraphics[width=70mm]{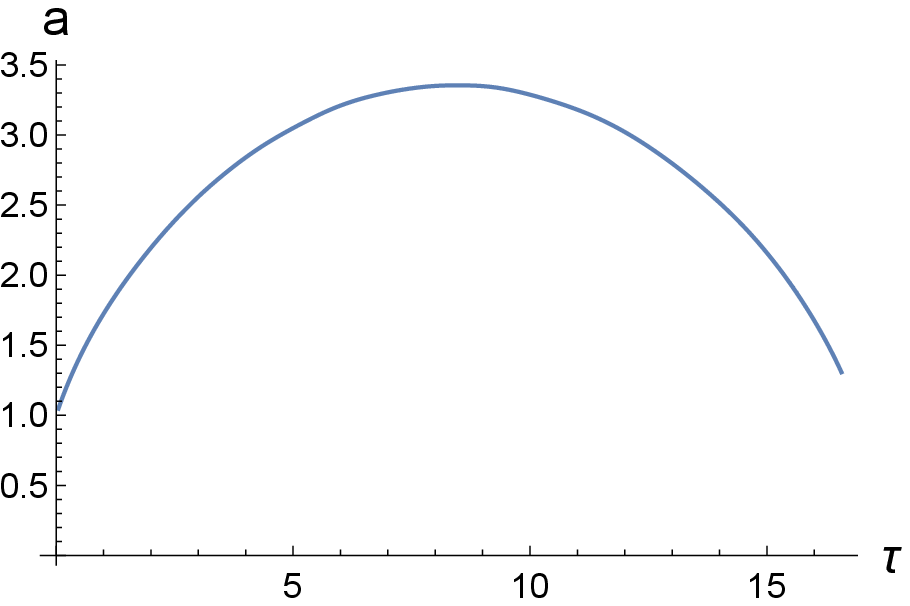} \hspace{8mm}
\includegraphics[width=70mm]{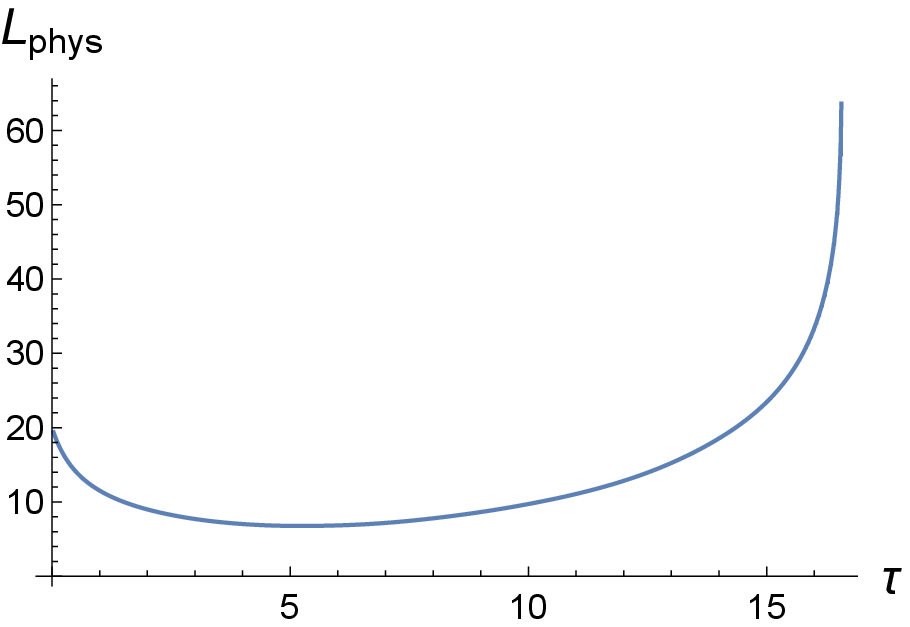} \\
\vspace{5mm}
\includegraphics[width=70mm]{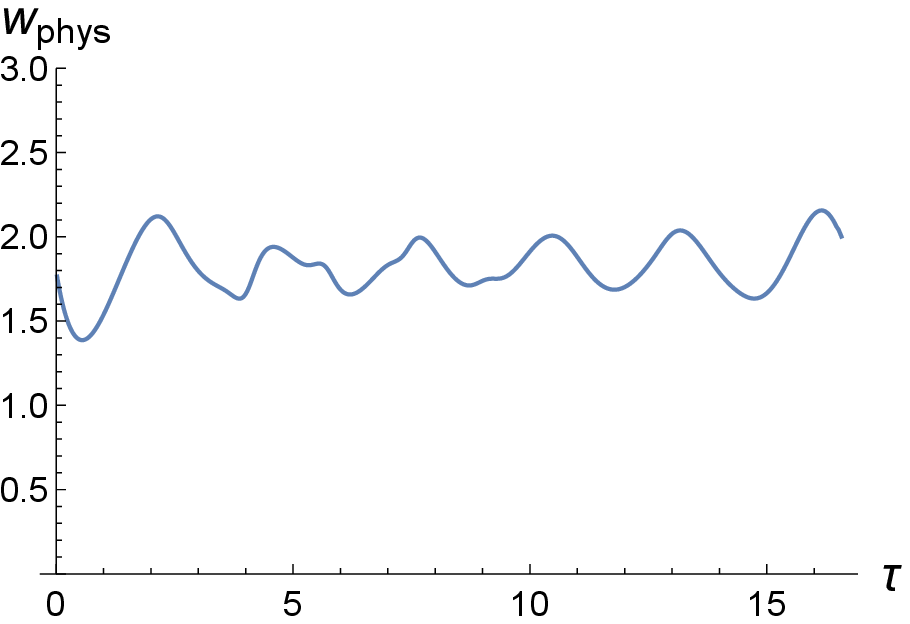} \hspace{8mm}
\includegraphics[width=70mm]{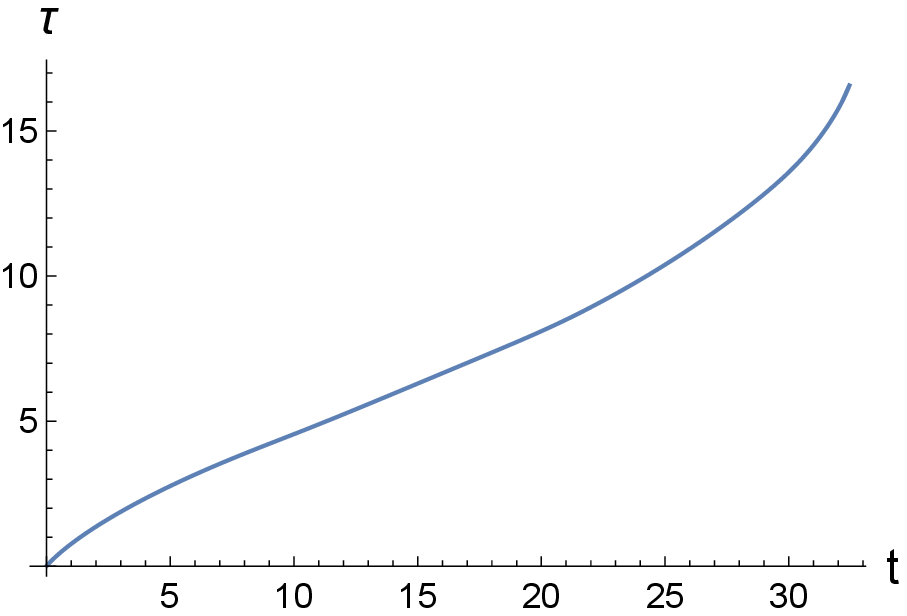}
\end{center}
\caption{The scale factor~$a$, the physical size of the extra dimension~$L_{\rm phys}$ 
and the physical width of the domain wall~$w_{\rm phys}$ as functions of $\tau$. 
The bottom-right plot shows the relation between $t$ and $\tau$. 
The parameters are chosen as (\ref{prm_choice}) and $V_{\rm min}=-0.05$.  }
\label{profile:aH05}
\end{figure}

Next we consider a case of $V_{\rm min}<0$. 
Figs.~\ref{profile:Phi05} and \ref{profile:A05} show the profiles of $\Phi$, 
$A$ and $B$ at each time for $V_{\rm min}=-0.05$ ($C_1=0.05$). 
Similar to the case of $V_{\rm min}\geq 0$, the scalar configuration once loses the kink profile, 
and then reconstructs the kink after some time. 
However, the kink configuration becomes wavy and seems unstable at $\tau\simeq 16.6$.  
At the time, both $A$ and $B$ have peaks at $z=5,15$, which are 
the middle points between the walls. 
This is in contrast to the case of $V_{\rm min}\geq 0$. 
More importantly, $B$ decreases at late times. 
This indicates that our 3D space will shrink (see Fig.~\ref{profile:aH05}). 

As we can see from Fig.~\ref{profile:aH05}, 
the size of the extra dimension~$L_{\rm phys}$ diverges at a finite value of $\tau$. 
So we cannot continue the numerical calculation beyond this time. 
This is before the scalar configuration reaches the step function profile. 
Beyond this time, the theory should be treated as 5D theory with non-compact extra dimension. 
The wall width~$w_{\rm phys}$ roughly stays constant during the evolution. 

\subsection{Case of $\bdm{\dot{B}(0,z)<0}$}
\begin{figure}[H]
\begin{center}
\includegraphics[width=70mm]{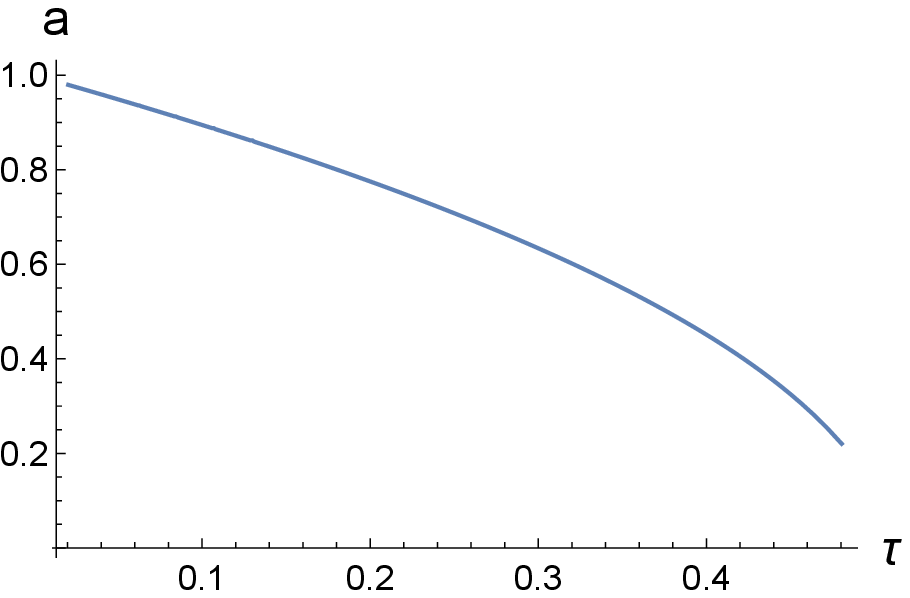} \hspace{8mm}
\includegraphics[width=70mm]{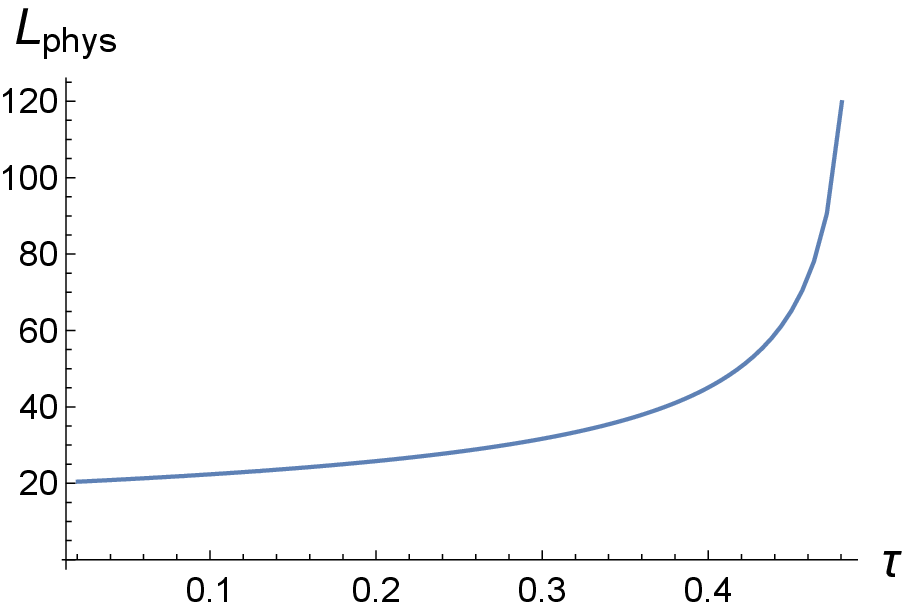} \\
\vspace{5mm}
\includegraphics[width=70mm]{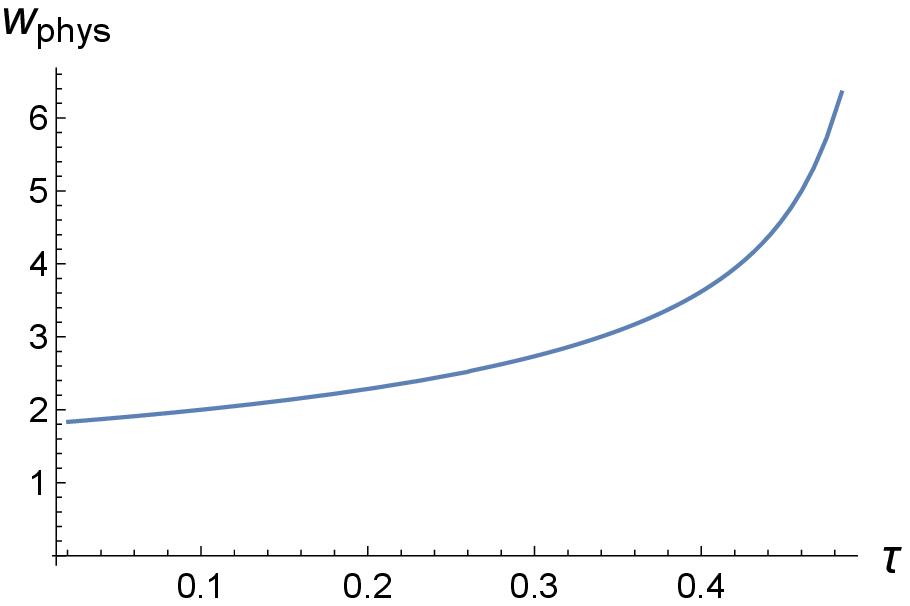} \hspace{8mm}
\includegraphics[width=70mm]{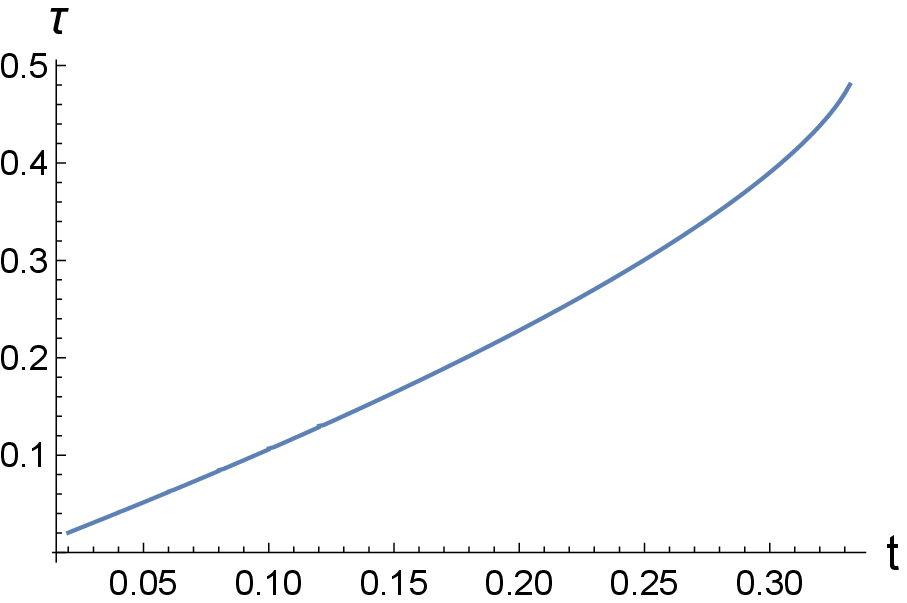} 
\end{center}
\caption{The scale factor~$a$, the physical size of the extra dimension~$L_{\rm phys}$ 
and the physical width of the domain wall~$w_{\rm phys}$ as functions of $\tau$. 
The bottom-right plot shows the relation between $t$ and $\tau$. 
The parameters are chosen as $V_{\rm min}=0$ and (\ref{prm_choice}) except for $b_0=-1$.  }
\label{profile:aH-b0minus}
\end{figure}

Before concluding, 
we also see the case that the initial value of $\dot{B}$ is negative, 
in order to see the dependence on the initial condition. 
The results are shown in Fig.~\ref{profile:aH-b0minus}. 
In this case, the scale factor monotonically decreases while the extra dimension diverges 
at a finite value of $\tau$. 
Our calculations fail at this time. 
These behaviors are similar to those in the case of Sec.~\ref{negV_0} at late times. 
In fact, $\dot{B}$ is negative at those times.  
However, the physical width of the domain wall~$w_{\rm phys}$ increases as the extra dimension expands 
in the current case. 

\section{Summary} \label{summary}
We have investigated the time evolution of the domain-wall configuration 
in the 5D gravitational theory with a compact extra dimension. 
In contrast to the case of non-gravitational theories, 
there is no static domain-wall solution in the gravitational theory. 
This is because the positive tensions of the domain walls prevent the warp factor 
from satisfying the periodic boundary condition along the extra dimension. 
Hence the domain-wall configuration evolves with time. 

In the case that the minimal value of the scalar potential~$V_{\rm min}$ is non-negative, 
both the 3D space and the extra dimension expand at late times. 
The former expands exponentially, while the latter does linearly with the cosmic time. 
The Hubble parameter for the 3D space approaches a positive value even in the case of $V_{\rm min}=0$. 
In the case of negative value of $V_{\rm min}$, 
the 3D space eventually shrinks while the extra dimension diverges at a finite cosmic time. 
Beyond that time, we cannot discuss the physics in our setup. 

In our calculations, we have assumed that $A$ and $B$ have constant profiles in the extra dimension 
at the initial time, just for simplicity.  
They will develop nontrivial profiles at late times. 
In the case of $V_{\rm min}\geq 0$, they will have peaks at the positions of the domain walls (Fig.~\ref{profile:A05}). 
This is similar to the Randall-Sundrum model~\cite{Randall:1999vf}, 
in which the 3-brane with a positive tension warps the ambient geometry 
and the warp factor has a peak at the brane. 
In the case of $V_{\rm min}<0$, on the other hand, $A$ will have peaks at the middle points between the walls. 
This indicates that, in the extra-dimensional direction, the region between the walls 
will expand faster than around the walls. 
As for the 3D space, $B$ will have mimima at the middle points (Fig.~\ref{profile:A05}). 
Thus, the non-compact 3D space directions will shrinks faster there 
compared to those at the wall positions. 
Recalling that the scale factor~$a(\tau)$ is mainly affected by the geometry around the walls 
due to the wave function of the observer (see (\ref{def:line_el}) and (\ref{def:a})), 
the 3D space in the bulk region shrinks faster than that shown in Fig.~\ref{profile:aH05}, 
and will collapse.  

Although Fig.~\ref{profile:Phi} shows that the scalar profile approaches the singular 
(periodic) step function, this just means that the ratio of the wall width to the size of the extra dimension
becomes small due to the expansion of the latter. 
In fact, the physical wall width~$w_{\rm phys}$, 
which determines the mass scale of the excited modes localized at the wall, 
does not decrease. 
It almost remains constant during the evolution. 

If we choose the initial condition such that the initial value of $\dot{B}$ is negative, 
the behavior of the configuration is similar to those at late time in the case of Sec.~\ref{negV_0}. 
Thus, our calculations fails at a finite time. 

In summary, the background configuration evolves without the collapse of the 3D space 
only when $V_{\rm min}$ is non-negative and $b_0$ is positive. 
Otherwise, the setup will be destabilized at a finite cosmic time. 
The extra dimension always expands at late times 
while whether the 3D space expands or shrinks depends on the sign of $V_{\rm min}$. 
The former property is related to the fact that the kink configuration feels the repulsive force 
from other kinks. 

For the purpose of constructing a realistic model, the extra dimension must be stabilized 
at some finite value. 
Inspired by the Goldberger-Wise mechanism~\cite{Goldberger:1999uk}, 
an extra 5D scalar field might be necessary. 
The introduction of the extra scalar that induces an attractive force between the kinks  
makes it possible to stabilize the extra dimension. 
Therefore, an extension of our analysis to a model that has multi scalar fields with 
nontrivial topological windings is an intriguing subject. 
We will discuss this issue in a subsequent paper.

\subsection*{Acknowledgements}
H.A. is supported by Institute for Advanced Theoretical and Experimental Physics, 
Waseda University.

\appendix

\section{Jacobi amplitude} \label{Ja}
The Jacobi amplitude~${\rm am}(u,k)$~\footnote{ 
This function is also denoted as ${\rm am}(u|m)$, where $m\equiv k^2$. 
}
is defined as an inverse function of
\be
 u(\vph) = \int_0^\vph\frac{d\tht}{\sqrt{1-k^2\sin^2\tht}} = F(\vph,k),  
\ee
where $F(\phi,k)$ is the incomplete elliptic integral of the first kind. 

\begin{figure}[t]
\begin{center}
\includegraphics[width=7cm]{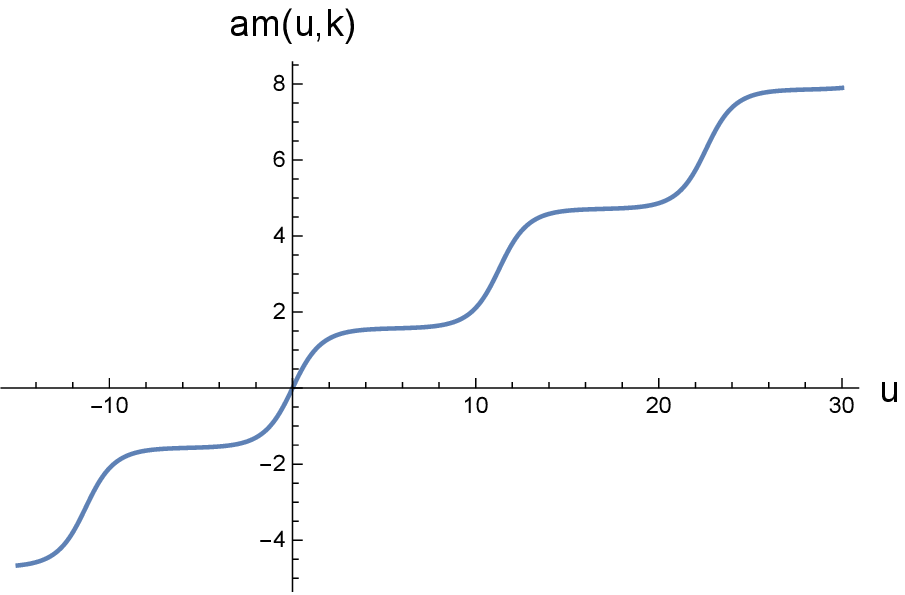} \hspace{5mm}
\includegraphics[width=7cm]{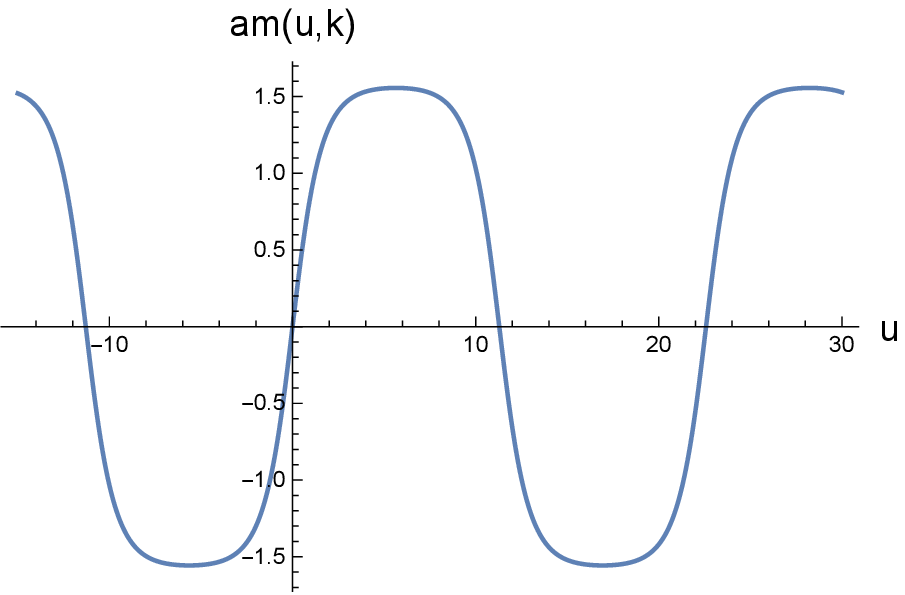}
\end{center}
\caption{The Jacobi amplitude~${\rm am}(u,k)$. 
The parameter~$k$ is chosen as $k=0.9999$ in the left plot, and $k=1.0001$ in the right plot. }
\label{profile:am}
\end{figure}

This function satisfies that
\be
 {\rm am}(0,k) = 0, \;\;\;\;\;
 {\rm am}(K(k),k) = \frac{\pi}{2}, 
\ee
where $K(k)$ is the complete elliptic integral of the first kind, which is given by
\be
 K(k) \equiv \int_0^{\pi/2}\frac{d\tht}{\sqrt{1-k^2\sin^2\tht}}. 
\ee
For $0<k<1$, ${\rm am}(u,k)$ is a monotonically increasing function that satisfies 
\be
 {\rm am}(u+2K(k),k) = {\rm am}(u,k)+\pi.  \label{shift:am}
\ee
(See the left plot in Fig.~\ref{profile:am}.) 
For $k>1$, ${\rm am}(u,k)$ is a periodic function with the period~$4F(\arcsin(1/k),k)$
(see the right plot in Fig.~\ref{profile:am}). 
In the limit of $k\to 1$, ${\rm am}(u,k)$ becomes the kink-like function, 
\be
 {\rm am}(u,1) = \arctan(\sinh u) = 2\arctan(e^u)-\frac{\pi}{2}. 
\ee

The derivative of ${\rm am}(u,k)$ is given by
\bea
 \frac{\der}{\der u}{\rm am}(u,k) \eql {\rm dn}(u,k), \nonumber\\
 \frac{\der^2}{\der u^2}{\rm am}(u,k) \eql \frac{\der}{\der u}{\rm dn}(u,k) = -k^2{\rm sn}(u,k){\rm cn}(u,k), 
 \label{der_am}
\eea
where 
\bea
 {\rm sn}(u,k) \defa \sin({\rm am}(u,k)), \nonumber\\
 {\rm cn}(u,k) \defa \cos({\rm am}(u,k)), \nonumber\\
 {\rm dn}(u,k) \defa \sqrt{1-k^2\sin^2({\rm am}(u,k))}, \;\;\;\;\;(\mbox{for $0\leq k\leq 1$})
 \label{def:sns}
\eea
are the Jacobi elliptic functions. 
For $k>1$, ${\rm dn}(u,k)$ is defined as
\be
 {\rm dn}(u,k) \equiv (-1)^l\sqrt{1-k^2\sin^2({\rm am}(u,k))}, 
\ee
for $(2l-1)u_0\leq u\leq (2l+1)u_0$ $(l\in{\mathbb Z})$, where~\footnote{
Note that ${\rm sn}(u_0,k)=1/k$, and ${\rm dn}(u_0,k)=0$. 
} 
\be
 u_0 \equiv F\brkt{\arcsin\brkt{\frac{1}{k}},k}. 
\ee

\section{Static limit of (\ref{t-EOM:1}) and (\ref{t-EOM:2})} \label{static_limit_EOM}
In the static limit, the background functions reduces as (\ref{static_limit}), and 
\bea
 \der_z\tl{\sgm} \toa e^\sgm\sgm, \;\;\;\;\;
 \der_z^2\tl{\sgm} \to e^{2\sgm}\brkt{\sgm''+\sgm^{\prime 2}}, \nonumber\\
 \der_z\tl{\Phi} \toa e^{\sgm}\Phi', \;\;\;\;\;
 \der_z^2\tl{\Phi} \to e^{2\sgm}\brkt{\Phi''+\sgm'\Phi'}, 
\eea
where the prime denotes the $y$-derivative 
($y$ is defined by (\ref{def:y})). 
Then, (\ref{t-EOM:1}) and (\ref{t-EOM:2}) become 
\bea
 &&\sgm''-2\sgm^{\prime 2}+\frac{\kp_5^2}{2}\brc{\Phi^{\prime 2}-\frac{2}{3}V(\Phi)} = 0, \nonumber\\
 &&\sgm''+4\sgm^{\prime 2}+\frac{2\kp_5^2}{3}V(\Phi) = 0, \nonumber\\
 &&\Phi''+4\sgm'\Phi'-V(\Phi) = 0,  \label{static:EOM:1}
\eea
with 
\bea
 0 \eql 0, \nonumber\\
 \sgm''+2\sgm^{\prime 2} \eql -\frac{\kp_5^2}{6}\brc{\Phi^{\prime 2}+2V(\Phi)}.  \label{static:EOM:2}
\eea

The last equation in (\ref{static:EOM:1}) is the same as the last one in (\ref{EOM:static}). 
From the first two equations in (\ref{static:EOM:1}), we have
\bea
 &&3\sgm''+\kp_5^2\Phi^{\prime 2} = 0, \nonumber\\
 &&6\sgm^{\prime 2}+\kp_5^2\brc{-\frac{1}{2}\Phi^{\prime 2}+V(\Phi)} = 0. 
\eea
These are equivalent with (\ref{EOM:redef}). 
The remaining equation in (\ref{static:EOM:2}) is equivalent with the first equation in (\ref{EOM:static}). 

Therefore, the static limit of (\ref{t-EOM:1}) and (\ref{t-EOM:2}) are reduced to (\ref{EOM:static}).


\end{document}